\documentclass[3p,10pt,sort&compress, colorlinks,linkcolor=blue, citecolor=blue, urlcolor=blue]{article}
\usepackage[utf8]{inputenc}

\usepackage{amssymb}
\usepackage{amsmath,mathrsfs}
\usepackage{empheq}

\usepackage[sc]{mathpazo}
\usepackage{bm}

\usepackage{color}

\usepackage{extarrows}

\usepackage{verbatim}

\usepackage{lineno}

\usepackage{siunitx}

\usepackage{stfloats}

\usepackage{bm}

\usepackage{graphicx}

\usepackage{caption}
\usepackage{subcaption}

\usepackage{multicol}

\usepackage{multirow}

\usepackage[version=4]{mhchem}

\usepackage[toc,page]{appendix}

\usepackage[super]{nth}

\usepackage{bm}

\usepackage{float}

\usepackage{threeparttable}

\usepackage{xcolor}

\usepackage{sectsty}
\usepackage[toc,page]{appendix}
\usepackage[colorlinks,linkcolor=blue,anchorcolor=green,citecolor=blue]{hyperref}

\newcommand{\secref}[1]{Sec.~\ref{#1}}

\newcommand{\figref}[1]{Fig.~\ref{#1}}
\newcommand{\subfigref}[2]{Fig.~\ref{#1}\ce{#2}}

\newcommand{\subssfigref}[3]{Figs.~\ref{#1}\ce{#2}-\ref{#1}\ce{#3}}
\newcommand{\tabref}[1]{Table~\ref{#1}}
\renewcommand{\eqref}[1]{Eq.~$($\ref{#1}$)$}
\newcommand{\eqsref}[2]{Eqs.~$($\ref{#1}$)$-$($\ref{#2}$)$}

\usepackage[sc]{mathpazo}

\def\diff{\mathrm{d}}

\newcommand*{\uW}[1]{\mathop{}\!\underline{W}_\mathrm{#1}}
\newcommand*{\ukappa}[1]{\mathop{}\!\underline{\kappa}_{#1}}
\newcommand*{\E}[1]{\mathop{}\!\times 10^{#1}}

\def\Tm{T_\mathrm{M}}
\def\T0{T_\mathrm{0}}

\def\stress{\boldsymbol{\sigma}}
\def\strain{\boldsymbol{\varepsilon}}
\def\stressdev{\mathbf{s}}

\def\vm{\sigma_\mathrm{e}}
\def\pe{p_\mathrm{e}}
\def\Hm{\mathsf{H}}

\def\ystress{\sigma_\mathrm{y}}

\def\barvm{\bar{\sigma}_\mathrm{e}}
\def\barpe{\bar{p}_\mathrm{e}}
\def\barT{\overline{T}}
\def\QQQ{\mathrm{Q3}}

\def\pbvm{\bar{\sigma}_\mathrm{e}^\mathrm{pb}}
\def\pbpe{\bar{p}_\mathrm{e}^\mathrm{pb}}

\def\subst{\mathrm{ss}}
\def\at{\mathrm{at}}
\def\sf{\mathrm{sf}}
\def\gb{\mathrm{gb}}

\def\lgb{l_\gb}

\def\Po{P}
\def\v{v}
\def\Pv{\Po\text{-}\v}
\def\vvec{\vec{v}}
\def\uvec{\vec{u}}
\def\nvec{\vec{n}}
\def\Cten{\mathbb{C}}

\def\DL{D_\mathrm{L}}

\def\cib{\mathrm{CI}_{95\%}}
\def\uv{U}

\def\pbvm{\barvm^\mathrm{P}}
\def\pbpe{\barpe^\mathrm{P}}

\def\fsvm{\barvm^\mathrm{S}}
\def\fspe{\barpe^\mathrm{S}}

\usepackage{cuted}%\stripsep 3pt plus 3pt minus 2pt
\usepackage{flushend}

\usepackage{enumerate}

\usepackage{authblk}

\providecommand{\keywords}[1]
{
	\small	
	\textbf{\textit{Keywords---}} #1
}

\begin{document}
	%opening
	
\title{Elasto-plastic residual stress analysis of selective laser sintered porous materials based on 3D-multilayer thermo-structural phase-field simulations}
% \title{3D-multilayer simulation of microstructure evolution and thermo-plastic behavior of porous materials by selective laser sintering}
	\author[1]{Yangyiwei Yang}
	\author[1]{Somnath Bharech}
  \author[1]{Nick Finger}
	\author[2,*]{Xiandong Zhou}
	% \ead{xdzhou@scu.edu.cn}
  \author[3]{Jörg Schröder}
	\author[1,*]{Bai-Xiang Xu}
	% \ead{xu@mfm.tu-darmstadt.de}

	\affil[1]{\small Mechanics of Functional Materials Division, Institute of Materials Science, Technische Universit\"at Darmstadt, Darmstadt 64287, Germany}
	
	\affil[2]{Failure Mechanics and Engineering Disaster Prevention Key Laboratory of Sichuan Province, College of Architecture and Environment, Sichuan University, Chengdu 610207, China}

        \affil[3]{Fakultät für Ingenieurwissenschaften, Abteilung Bauwissenschaften, Institut für Mechanik, Universität Duisburg-Essen, Essen 45141, Germany}

\affil[*]{Corresponding authors: \url{xu@mfm.tu-darmstadt.de} (Bai-Xiang Xu), \url{xdzhou@scu.edu.cn} (Xiandong Zhou)}

\date{}
\maketitle
\renewcommand\Authands{ and }
		
\begin{abstract}
Residual stress and plastic strain in additive manufactured materials can exhibit significant microscopic variation at the powder scale, profoundly influencing the overall properties of printed components. This variation depends on processing parameters and stems from multiple factors, including differences in powder bed morphology, non-uniform thermo-structural profiles, and inter-layer fusion.
In this research, we propose a powder-resolved multilayer multiphysics simulation scheme tailored for porous materials through the process of selective laser sintering. This approach seamlessly integrates finite element method (FEM)-based non-isothermal phase-field simulation with thermo-elasto-plastic simulation, incorporating temperature- and phase-dependent material properties. 
The outcome of this investigation includes a detailed depiction of the mesoscopic evolution of stress and plastic strain within a transient thermo-microstructure, evaluated across a spectrum of beam power and scan speed parameters. Simulation results further reveal the underlying mechanisms. For instance, stress concentration primarily occurs at the necking region of partially melted particles and the junctions between different layers, resulting in the accumulation of plastic strain and residual stress, ultimately leading to structural distortion in the materials. Based on the simulation data, phenomenological relation regarding porosity/densification control by the beam energy input was examined along with the comparison to experimental results. Regression models were also proposed to describe the dependency of the residual stress and the plastic strain on the beam energy input. 
\end{abstract}
		

%\section{Introduction}\label{sec:intro}
%\subsection{beam powder bed fusion}
\section{Introduction}

% \subsection{LAM, SLS and porous material (YYang)}

% \begin{itemize}
%  \item Possible application of SLS for manufacturing porous composite exhibiting flexoelectric effect to make sensors and energy harvesters. (Check research works from Prof. Lihua Shao and Prof. JainXiang Wang)

%  \rsb{\item High porosity, in the range 25-40\% have been observed in SLS produced H13 tool steel parts \cite{ibraheem2002thermal}}
% \end{itemize}

% General AM Introduction, from Nick, needed to be integrated

In the recent decade, additive manufacturing (AM) emerged from a niche technology to a widely applicable means of production. Among a variety of AM techniques, powder bed fusion (PBF) stands out as a strong technique for producing intricate geometries while maintaining good structural integrity and superior material properties \cite{gu2021material, Ladani2021Metals, Zhang2017JournalofMaterialsEngineeringandPerformance, panwisawas2020metal, korner2020modeling}. In PBF, layers of powder get fused by a beam one after another to build up three-dimensional (3D) geometries. Compared to conventional ones, this manufacturing process offers several advantages, such as great design flexibility, reduction in waste, enabling rapid tooling and decreasing production cycles.

Selective laser sintering (SLS) is one prominent example of PBF, which utilizes the tuned incident beam (mostly continuous laser scan or laser pulse) to bind the free powders layer by layer. Due to its well-controlled powder bed temperature compared to other PBF techniques like selective laser melting (SLM), where the significant melting phenomenon can be captured, SLS enables only partial melting of particles and produces samples with relatively high porosity 
\cite{kim2020fabrication, yang2019npj, gu2012laser, Li2018, li2010316l}. In this regard, SLS has great potential in applications that require complex geometries with high porosity. For instance, SLS is applicable in manufacturing porous components for medication applications, especially medical scaffolds, and artificial bones \cite{shuai2019construction, duan2010three, eshraghi2010mechanical}. 
It is feasible in manufacturing functional materials with a low processing temperature, such as ferromagnetic materials \cite{mapley2019selective, wendhausen2017,  sing2017direct, jhong2016fabricating, huang2019}. 
By precisely controlling the geometry and structure of the printed material, it may also be possible to create the desired strain gradients and electric polarization necessary to generate the flexoelectric effect \cite{Zhang2022NanoEnergy, Zhang2021JournaloftheMechanicsandPhysicsofSolids, Yan2023ScienceAdvances}.

Residual stress has been a critical issue since it can affect mechanical properties, dimensional accuracy, corrosion resistance, crack growth resistance, and performance of AM parts~\cite{chen2022review}. %In general, there are three types of residual stresses classified by the length scale. Type I residual stresses arise from non-uniform plastic deformation at the part scale, which results in large deformation of the part. Type II residual stress is microscale stress acting at the individual grain scale. Type III residual stress is nanoscale stress due to coherency and dislocation. Type II and Type III residual stresses have very limited effect on the material’s mechanical properties~\cite{mercelis2006residual}. Therefore, aiming to study the formation of global residual stress arising from non-uniform plastic deformation in the porous material manufactured through the SLS process, only type I residual stress is considered in this work.
There are already investigations focusing on the spatial distribution of the residual stress within the AM-built parts, which were carried out by experimental measurements and theoretical estimations \cite{mercelis2006residual,roberts2010numerical, pant2021simplified, mukherjee2017improved}. Mercelis et al. explained the origins of residual stress in parts produced by AM. They experimentally measured the residual stress distribution in a selective laser sintering produced part and compared it with analytical and numerical solutions \cite{mercelis2006residual}. Pant et al. employed a layer-by-layer finite element approach to predict the residual stress and validated the model using measurements from neutron diffraction \cite{pant2021simplified}. Ibraheem et al. predicted the thermal profile and residual stress of SLS processed H13 tool steel using a finite element model. High porosity was observed in the range 25-40\%, yet a homogenized powder bed was utilized to simulate the thermal evolution and subsequently the residual stress \cite{ibraheem2002thermal}.

The calculation of residual stress in literature is often performed in two sequential steps: first the transient temperature is simulated in the entire domain and then the thermal history is used to calculate the stress and strain evolution in thermo-mechanical simulations. The accuracy of these calculations critically relies on the quality of the thermal history. 
The temperature gradient mechanism (TGM) model and the cool-down stage (CDS) model are two commonly used models to explain the development of plastic strain and residual stress formation mechanisms~\cite{mercelis2006residual, simson2017, takezawa2022}. In the TGM model, the deformation of the material in the molten pool/fusion zone is restrained by surrounding materials during both the heating and cooling stages due to a large temperature gradient inside and outside the overheated region (where the on-site temperature is beyond the melting point). In the heating stage, the plastic strain is developed in the surrounding material due to an expansion of the heated material. As the heated material cools down, it starts to shrink, which is, however, counteracted by the plastically deformed surrounding material. Thus, a tensile residual stress develops. The CDS model, on the other hand, was proposed to elucidate the residual stress due to the layerwise features of the AM process and the AM-manufactured parts. This occurs because, in both the heating and cooling stages, the deformation of the upper-layers is restrained by the fused lower-layers and/or substrate. Meanwhile, melted material in the lower-layers will undergo a remelting and re-solidification cycle. Both factors result in tensile residual stress in the top layer due to the shrinkage of the material during the cooling stage. 
These two models provide phenomenological aspects by employing idealized homogeneous layers for the understanding of the formation mechanism of residual stresses. They have been widely adopted in numerical analysis of residual stress at single layer scale models~\cite{yi2019computational, hussein2013finite,parry2016understanding,vastola2016controlling} and part-scale multilayer models~\cite{chiumenti2017numerical,ganeriwala2019evaluation,carraturo2020modeling}. 

In these aforementioned works, homogeneous layers are used for analyses of the thermal history and the mechanical response in an AM-manufactured part. However, due to the complex morphology of the powder bed on the powder scale and the resultant inhomogeneity of the local thermal properties, the thermal profile is subjected to a high level of non-uniformity as well. This implies heterogeneity of the thermal stress, the plastic deformation and the residual stress on the powder scale. In fact, mechanical properties of AM-built parts, such as fracture strength and hardening behavior, are significantly influenced by the local defects and local stress \cite{chen2019a, bartlett2019}. 
The diversity in the powder bed structure and the packing density introduces thermal heterogeneity in the form of mesoscopic high-gradient temperature profile and asynchronous on-site thermal history. Such variability leads to varying degrees of thermal expansion and, therefore, thermal stresses. Additionally, the presence of stochastic inter-particle voids and lack-of-fusion pores contribute to evolving disparities in thermo-mechanical properties \cite{yang2019npj, zhou20213d, yang2023npj}. The thermal gradient across the powder particles is very sharp as per the modeling results by Panwisawas et al. \cite{panwisawas2015role}, which is in agreement with experimental fundings \cite{arisoy2019modeling} and is also recaptured in our former numerical works \cite{yang2019npj, zhou20213d, yang2023npj}. Furthermore, the layerwisely build-up process results in local interaction between the newly deposited and previously deposited layers with high surface roughness. Thus, the residual stress formation on the powder scale should be revisited for the SLS process. 

For this purpose, the powder bed morphology, the heat transfer and the porous microstructure evolution during the SLS process should be primarily considered. One promising approach is the phase-field simulation. For instance, phase-field simulations of the AM process can reveal the in-process microstructure evolution, providing insights into key features such as porosity, surface morphology, temperature profile, and geometry evolution~\cite{Zhang2022Coatings, yang2019npj}. Based on a non-isothermal phase-field multilayer simulation results of selective laser sintered porous structures, thermo-elastic simulations and homogenization of the elastic properties have been investigated in our previous work~\cite{zhou20213d}. In combination with the single layer phase-field thermo-microstructural simulations, the first powder-resolved elasto-plastic simulations of the residual stress in a SLS-processed magnetic alloy has also been performed~\cite{yang2023npj}. 
Results show that during the cooling stage, the partially melted and interconnected particles result in plastic deformation due to the shrinkage of the fusion zone, but mostly at the necking region because of stress concentration in porous microstructures. By combining the thermo-structural phase-field method and the thermo-mechanical calculations, this work demonstrates the promising capability of the multiphysics simulation scheme for the local residual stress analysis of AM-built materials on the powder scale. 
In the current article, we extend this multiphysics simulation scheme for elasto-plastic multilayer SLS process, providing systematic simulations of the local plastic deformation and the residual stress. These allow us to reveal the influential aspects related to the printing process and the powder bed.

\section{Results}

% {
% some discussion points
% \begin{itemize}
% \item different thermal condition/history on the microstructure along with morphologies (pores/substance) $\rightarrow$ different stress/strain states
% \item relation between stress/strain states and multilayers 
% \end{itemize}}

\begin{figure}[!h]
	\centering
 	\includegraphics[width=13cm]{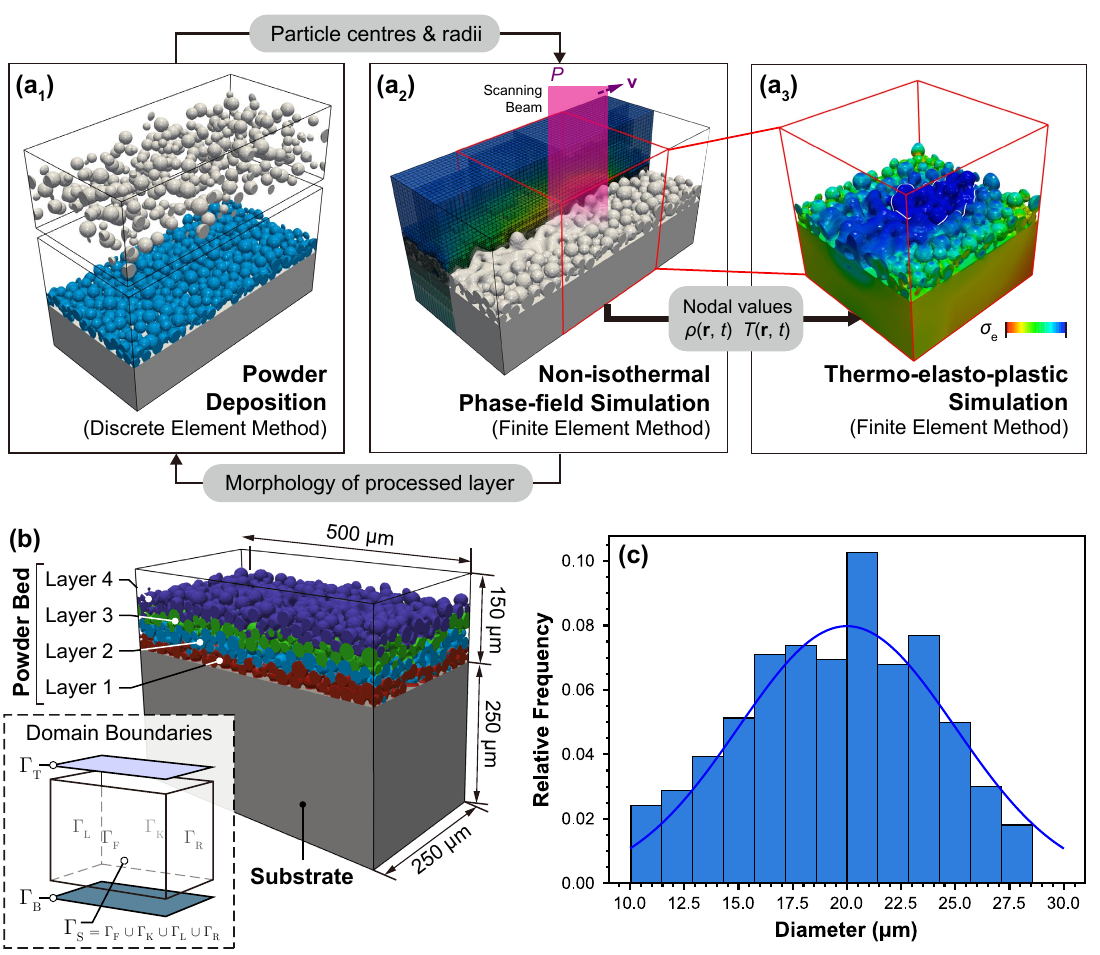}
	\caption{\textbf{3D-multilayer simulation scheme for selective laser sintering}. (a) The workflow of multilayer selective laser sintering (SLS) simulation, incl. the layerwise powder deposition, the non-isothermal phase-field simulation and subsequent thermo-elasto-plastic calculation. (b) Prime simulation domain with four deposited-processed powder layers. Inset: domain boundaries and their denotations. (c) Powder size distribution of the common first layer.} 
 %(a$_{1}$) Powder deposition in the simulation domain adhering to the powder size and distribution characteristics as mentioned in Sec. $\ref{sec:para}$.} }
	\label{fig:wf}
\end{figure}

In this work, the simulation workflow is arranged in three stages, as shown in \figref{fig:wf}a. The powders were first deposited into a prime simulation domain based on the discrete element method (DEM) under a given gravitational force. Then, non-isothermal phase-field simulations were performed for the coupled thermo-structural evolution during the SLS process. Upon completion of a layer, the resulting microstructure is voxelized and re-imported into the DEM program for the deposition of the next powder layer until reaching the desired layer number or height. Meanwhile, the subsequent thermo-elasto-plastic calculations were performed to investigate the development of plastic strain and stress under the quasi-static thermo-structures by mapping the nodal values of the order parameters (OP, indicating the chronological-spatial distribution of the phases) and temperature to a subdomain. It is worth noting that the thermo-elasto-plastic calculation of the next layer was restarted from the former one with continuous history of the residual stress and accumulated plastic strain, though the OPs and temperature were reiterated in the prime domain. As it is usually done in the literature, we assume thereby that heat transfer is only strongly coupled with microstructure evolution (driven by diffusion and underlying grain growth), but mechanics impose negligible effects on both heat transder and microstructure evolution. 

{SLS processes with a constant beam spot diameter $\DL$, and various combinations of beam power $\Po$ as well as scan speed $v$ were simulated. The processing window $\Po\in[15,~30]~\si{W}$ and $v\in[75,~150]~\si{mm~s^{-1}}$ was set in order to compare it with previous studies \cite{yang2019npj, zhou20213d}. $\DL=D_\mathrm{FWE2}=200~\si{\micro\meter}$ (FWE2 represents the full-width at $1/e^2$) was adopted as the nominal diameter of the beam spot, within which around $86.5\%$ of the power is concentrated. The full-width at half maximum intensity (FWHM) then equals to $0.588D_\mathrm{FWE2}=117.6~\si{\micro\meter}$, characterizing $50\%$ power concentration within the spot. The powder bed is preheated with a preheating temperature $\T0=0.4\Tm = 680~\si{K}$, which was embodied by temperature initial condition (IC) and boundary condition (BC). Both powder and substrate are SS316L. Powders are firstly deposited into a $250\times500\times400$ \si{\micro m^3} prime simulation domain under a given gravitational force based on the discrete element method (DEM). This domain contains the free space of $150~\si{\micro m}$ height for positioning powders and a substrate of $250~\si{\micro m}$ thickness. In-total four layers of SLS processes were performed for each pair of processing parameters (i.e., $\Po$ and $v$, hereinafter as $\Pv$ pair). Each newly deposited layer maintains the same powder size distribution and a volume fraction of 12\% with respect to the free space of $250\times500\times150$ \si{\micro m^3}. This guarantees each deposited layer sufficiently holds one to two particles in thickness, as illustrated in \figref{fig:wf}b. The powder size follows Gaussian distribution with a mean diameter of $\mu_d = 20~\si{\micro m}$, a standard deviation $\varsigma_d = 5~\si{\micro m}$, and a cut-off bandwidth $d_\mathrm{cut} = 10~\si{\micro m}$. \figref{fig:wf}c presents the powder size distribution of the first layer (common for all $\Pv$ combinations). After each layer's beam scan (i.e., the beam leaves the prime domain), the simulation continues for twice the duration of the scanning period to emulate a natural cooling stage.} Details regarding the process as well as thermo-elasto-plastic modeling, simulation setup, and parameterization are explicitly given in the \hyperref[sec:method]{\textit{Methods}} section.

\subsection{Thermo-structural evolution during multilayer SLS}\label{sec:T}

\begin{figure}[!h]
	\centering
 	\includegraphics[width=18cm]{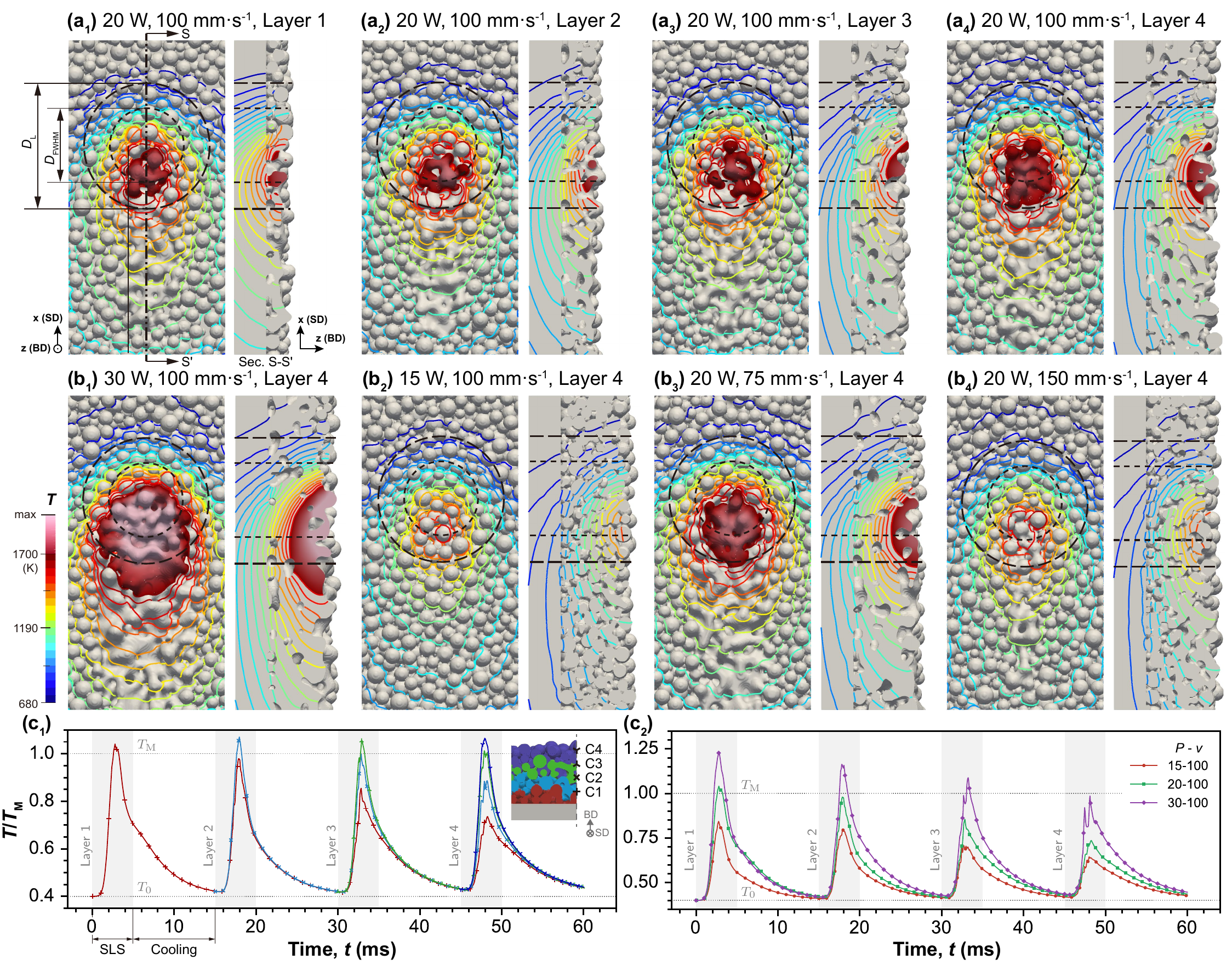}
	\caption{\textbf{Simulated thermo-structural evolution during four-layer SLS single scan}. (a$_1$)-(a$_4$) Transient thermo-structural profiles under a
 %beam power 20W and scan speed 100mm s$^{-1}$ 
  beam power $20 ~\si{W}$ and a scan speed $100 ~\si{mm~s^{-1}}$ with the beam spot consistently positioned across various layers. The beam spot size characterized by $\DL$ and $D_\mathrm{FWHM}$ is indicated.
  (b$_1$)-(b$_4$) Transient thermo-structural profiles of the \nth{4} layer under various beam power and scan speed with the beam spot consistently positioned. The overheated regions ($T\ge\Tm$) are illustrated by a continuous color map. (c$_1$)-(c$_2$) Temperature history of the selected surface points across layers at various beam power with scan speed maintained at 100 \si{mm~s^{-1}}. Single-layer SLS (shaded sections) and cooling phase are also denoted. Inset: the location of the points. }
	\label{fig:T}
\end{figure}

\figref{fig:T}a$_1$-a$_4$ present the transient thermo-structural profiles for a single scan of $\Po = 20 ~\si{W}$ and $\v = 100 ~\si{mm~s^{-1}}$ where the beam spot is consistently positioned along scan direction (SD) across various building layers. 
% The overheated region where $T > \Tm$ is depicted with a continuous colormap, while the regions with $T \le \Tm$ are drawn with temperature isotherms. 
In the overheated region ($T \ge \Tm$), particles can undergo complete or partial melting. This prompts molten materials to flow from convex to concave points, making the fusion of particles possible and, therefore, forging the fusion zone. In regions where the temperature remains below the melting point ($T < \Tm$), however, the temperature is still high enough to cause diffusion as the diffusivity grows exponentially w.r.t. temperature. This is evidenced by the necking phenomenon among neighboring particles. 
Temperature profiles demonstrate a strong dependence on the local morphology. Notably, the concentrated temperature isolines can be observed around the neck region among particles, indicating an increased temperature gradient (up to around $50$-$100~\si{K~\micro m^{-1}}$, comparing to $1~\si{K~\micro m^{-1}}$ in the densified region). It is worth noting that such thermal inhomogeneity induced by stochastic transient morphology can hardly be resolved by the simulation works employing homogenized powder bed \cite{takezawa2022, yang2022validated, mukherjee2017improved}.

The heat dissipation in the powder bed also changes considerably as upper-layers are continuously built and processed. This can be noticed by a varying shape and significance of the overheated region across layers. As a comparison, the overheated region at \nth{4} layer (\figref{fig:T}a$_1$) is greater in comparison to the initial layer (\figref{fig:T}a$_4$). This can simply be reasoned by increasing porosity in fused lower-layers that block the heat dissipation. At the initial layer, this dissipation is affected the least as there is no lower-layer but substrate. 
We further probed the temperature history at surface points located at the center of the scan path, as shown in \figref{fig:T}c$_1$, where the recurring peaks in every single-layer SLS stage (shaded section) present the thermal cycle. The first peaks of distinctive cycles (probed on different surface points) nearly match. In an identical cycle (e.g., C1), the peaks gradually decrease as the SLS processes advance. Once the spot leaves the probing point, the temperature drop is rapid at first (mostly due to the heat conduction driven by the high-gradient temperature profile) and then becomes moderate (due to heat convection and radiation) as the cooling stage begins, when the heat convection and radiation are effective. This steep temperature drop gradually disappears at points from fused lower-layers as the upper-layers are continuously built, comparing the thermal cycles at point C1. Meanwhile, the temperature drop at distinct points gradually coincides as the cooling stage proceeds, comparing the thermal cycles at points C1-C4.

The multilayer thermo-structural evolution is also significantly affected by processing parameters, as the profiles of varying beam powers and scan speeds presented in \figref{fig:T}b$_1$-b$_4$. Increasing the beam power and/or decreasing the scanning speed is observed to intensify heat accumulation at the beam spot, resulting in an enlargement of the overheated region. When $\v$ is held constant at $100~\si{mm~s^{-1}}$, decreasing $\Po$ from $30 ~\si{W}$ (\figref{fig:T}\ce{b1}) to $15 ~\si{W}$ (\figref{fig:T}\ce{b2}) results in a less pronounced overheated region. On the other hand, maintaining a constant $\Po = 20~\si{W}$ and increasing $v$ from $75~\si{mm ~s^{-1}}$ (\figref{fig:T}\ce{b3}) to $150~\si{mm~s^{-1}}$ (\figref{fig:T}\ce{b4}) leads to a reduction in the overheated region as well. It is also evident that increasing $\Po$ and/or decreasing $\v$ produces less porosity in fused lower-layers, which may further change the thermal conditions and improve the heat dissipation. \figref{fig:T}c$_2$ presents the probed thermal cycle of point C1 under various $\Po$ with $\v=100~\si{mm~s^{-1}}$ maintained. One can tell that the temperature of the case $\Po=30~\mathrm{W}$ quickly drops from a higher peak to the value coincided with one of the case $\Po=20~\mathrm{W}$ at the end of scan duration, implying an improved heat dissipation. It should also be noted that surface point C1, located at the initial layer, experienced three peaks of temperature that are beyond $\Tm$ due to the extended depth of the fusion zone at higher beam power. In Sec. \ref{sec:pr-poro} we will continue the discussion regarding the relation between porosity and processing parameters.

% {(morphology change may also incite absorptivity change, add later, short paragraph)}

\subsection{Stress and plastic strain evolution during multilayer SLS process}

\begin{figure}[!h]
	\centering
 	\includegraphics[width=18cm]{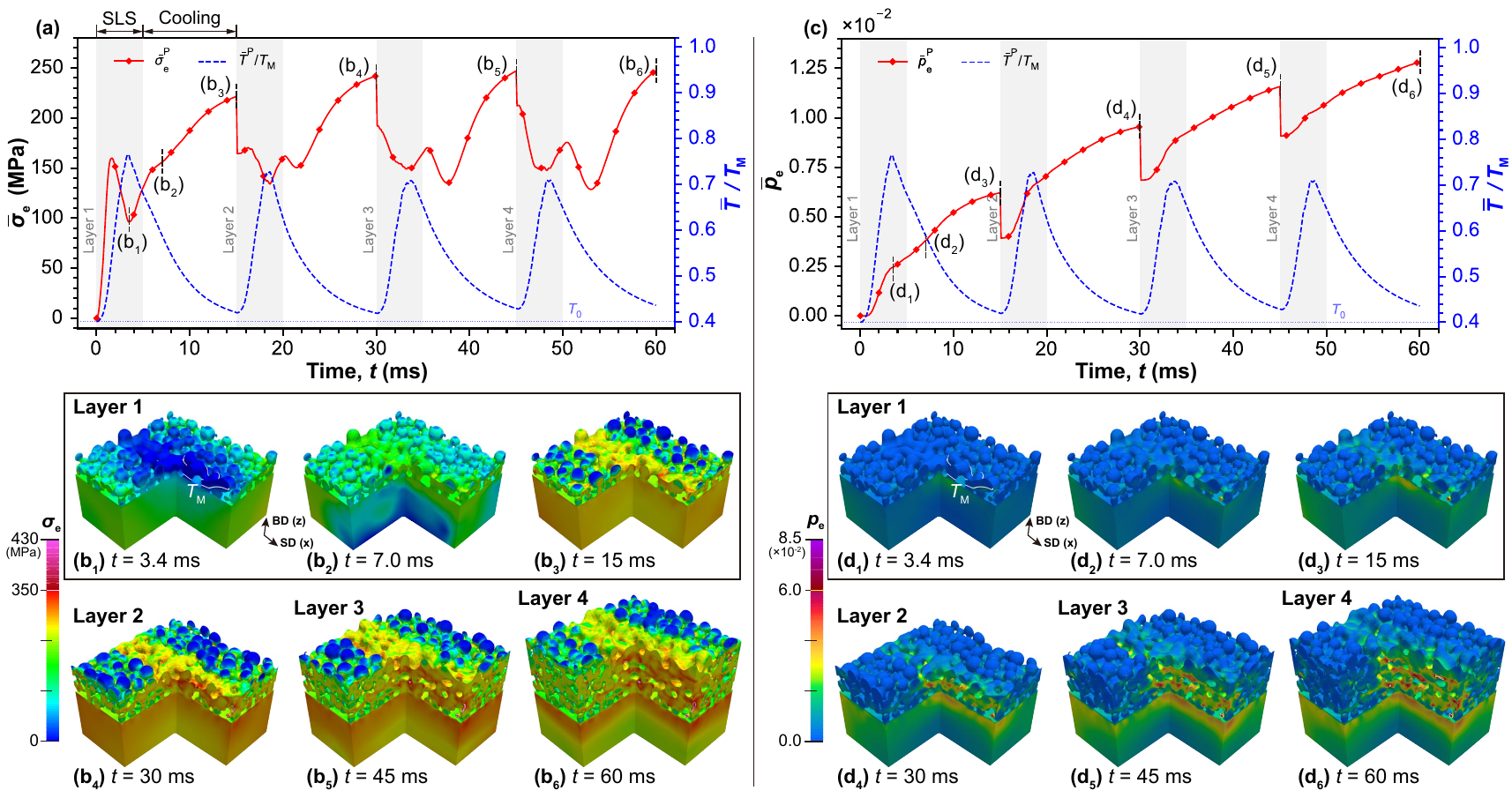}
	\caption{\textbf{Development of average stress and plastic strain during four-layer SLS single scan}. (a) Calculated von Mises stress ($\barvm^\mathrm{P}$) and temperature ($\barT^\mathrm{P}$) in powder-bed
average vs. time with the profile of $\vm$ at the denoted states shown in (\ce{b1})–(\ce{b6}). (c) Calculated effective plastic strain in powder-bed average ($\barpe^\mathrm{P}$) vs.
time with the profile of $\pe$ at the denoted states shown in (\ce{d1})–(\ce{d6}).}
	\label{fig:tep-avg}
\end{figure}

\begin{figure}[!h]
	\centering
 	\includegraphics[width=18cm]{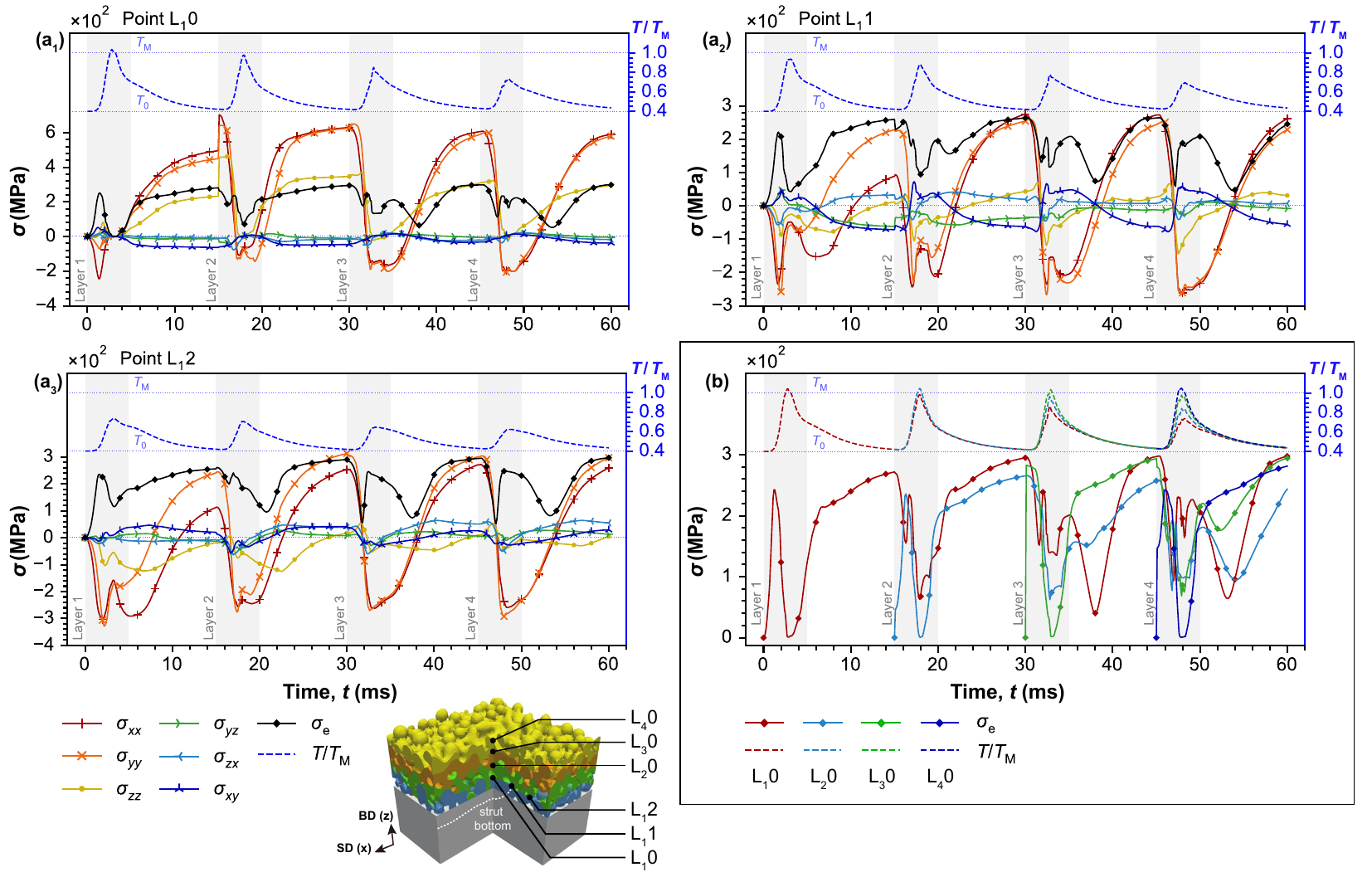}
	\caption{\textbf{Transient stress components at varying coordinates during four-layer SLS single scan}. Probed history of (\ce{a1})-(\ce{a3}) stress components at surface points L$_1 0$, L$_1 1$ and L$_1 2$, and (b) von Mises stress $\vm$ at surface points L$_1 0$, L$_2 0$, L$_3 0$ and L$_4 0$ across layers. Inset: Location of the selected points. Fusion zone boundary (FZB) of the initial layer is also denoted.}
	\label{fig:sig}
\end{figure}

% \subsubsection{Evolution of the average stress/plastic strain}
To analyze the overall stress and plastic strain evolution during SLS, the average quantities (incl. von Mises stress $\vm$, effective plastic strain $\pe$, and temperature $T$) in the powder bed are defined as
\begin{equation}
 \barvm^\mathrm{P}=\frac{\int_{\Omega'} \rho \vm\diff \Omega}{\int_{\Omega'} \rho\diff \Omega}, \quad\barpe^\mathrm{P}=\frac{\int_{\Omega'}^\mathrm{P} \rho \pe\diff \Omega}{\int_{\Omega'} \rho\diff \Omega},\quad\barT^\mathrm{P}=\frac{\int_{\Omega'}^\mathrm{P} \rho T\diff \Omega}{\int_{\Omega'} \rho \diff\Omega},
 \label{eq:pbavg}
\end{equation}
where $\Omega'$ is the volume of the simulation domain without the substrate meshes, and $\rho$ is the OP indicating the substance with $\rho=1$. They are hereinafter termed as PB-averaged quantities. \figref{fig:tep-avg}a presents that $\pbvm$ develops as the SLS process advances on distinctive layers. When the beam spot enters the domain, accompanied by the appearance of the overheated region, $\barvm^\mathrm{P}$ begins to decrease while $\barT^\mathrm{P}$ continues to rise to its peak. This is attributed to losing structural integrity caused by full/partial melting within the overheated region, resulting in zero stress as shown in \figref{fig:tep-avg}\ce{b1}. The areas surrounding the overheated region also exhibit relatively low stress due to a significant reduction in stiffness at elevated temperatures. Stress accumulates faster around concave features, such as surface depressions and particle sintering necks, compared with the stress around convex features as well as unfused powders away from the fusion zone. This is due to localized temperature gradients, as explained in Sec. \ref{sec:T}. As cooling progresses, stress decreases in convex features and unfused powders, yet remains concentrated around concave features (see Figs. \ref{fig:tep-avg}\ce{b2}-\ref{fig:tep-avg}\ce{b3}). The stress at the end of each cooling stage (Figs. \ref{fig:tep-avg}\ce{b3}-\ref{fig:tep-avg}\ce{b6}) serves as the residual stress of corresponding processed layers. After the deposition of new layers, $\barvm^\mathrm{P}$ has an instant drop due to the addition of stress-free substances (powders), then a new cycle begins. {It is worth noting that the peak values of the stress cycle (which are also residual stresses of each processed layer) are almost identical, implying that each single-layer SLS generates nearly the same amount of residual stress by average in the powder bed. For upper-layers, $\vm$ experiences an additional reduction in the cooling stage, which may be attributed to the more delicate variations in the on-site stress components and will be discussed in \figref{fig:sig}.}

The accumulated plastic strain $\pe$, on the other hand, presents an overall increasing tendency vs time during the single-layer SLS and cooling stage, as shown in \figref{fig:tep-avg}c. This distinguishes the plastic strain history from the stress cycle, which suffers a reduction during SLS due to loss of structural integrity. The continuous accumulation of $\pe$ leads to a locally concentrated profile in the vicinity of the melt zone (see \figref{fig:tep-avg}\ce{d2}-\ref{fig:tep-avg}\ce{d3}). The high temperature gradient at the front and bottom of the overheated region during single layer SLS also results in the localized $\pe$ surrounding the fusion zone (see \figref{fig:tep-avg}\ce{d3}-\ref{fig:tep-avg}\ce{d6}), where the existing high thermal stress locally activates the plastic deformation. A similar explanation applies to the concentrated $\pe$ around pores and concave features such as sintering necks near the fusion zone. In contrast, unfused powders and the substrate away from the melt zone show minimal $\pe$, since the thermal stress at these locations is insufficient to induce material plastification due to lower local temperatures. 

We probed the history of the stress components at six points, where L$_1 0$-L$_4 0$ locates at the center of \nth{1}-\nth{4} layers, L$_1 1$ locates at the boundary of the fusion zone, and L$_1 2$ locates outside the fusion zone in the \nth{1} layer, as shown in the inset of \figref{fig:sig}.
The results mainly show the difference of the stress field inside and outside the fusion zone. 
% For the point L$_10$, it is inside the fusion zone during the SLS process of the \nth{1} layer. 
Since the temperature cools down from above the melting temperature to pre-heating temperature, the rise of the elasticity leads to a significant increase of $\vm$ on point L$_10$ for temperature below $0.65T_M$. As a comparison, the increament of the $\vm$ becomes smaller for the point L$_11$ and the point L$_12$, as shown in \subssfigref{fig:sig}{a1}{a3}. It can also be seen that $\vm$ for the mentioned three points (L$_10$-L$_12$) during SLS of \nth{2}- and \nth{3} layers are very similar when they are all outside the fusion zone. Similar results can be observed for the development of normal stress components $\sigma_{ii}$ ($i=x,~y,~z$). The main differences in normal stress for the three points are mostly for the SLS process of the \nth{1} layer. $\sigma_{ii}$ are tensile in the fusion zone while compressive outside the fusion zone. This is due to the thermal stress in the fusion zone being tensile while it is compressive outside the fusion zone. Curves for the \nth{2} and \nth{3} layers are similar because the three points are all outside the fusion zone. Shear stress components $\sigma_{ij}$ ($i,j=x,~y,~z$; $i\neq j$) remain small inside and outside the fusion zone. The variation of the shear stress is comparably more visible at the boundary of the fusion zone (L$_1$1), as shown in \figref{fig:sig}\ce{a2}. This is due to a large gradient of the thermal strain and material mechanical properties across the boundary of the fusion zone. With a larger variation of the shear stress components, variation of $\vm$ at the highest temperature during SLS process of each layer is also more visible. The saturation of the stress field can be observed after the \nth{2} layer. It means that the upper layer has limited influence on the plastic deformation of previous layers, which experience lower temperatures and lower temperature gradients outside the fusion zone. 

For points in the layer-wise direction (L$_10$ to L$_40$ at the boundary of each layer), the results in \figref{fig:sig}b show a direct comparison between the development of the stress field of a point in the microstructure during SLS process of different layers. It is worth noting that the development of the $\vm$ of each point is very similar at the beginning. The saturation of the $\vm$ for each point can be observed in the layer-wise direction when the points are outside the fusion zone. However, variation of the $\vm$ is much more pronounced when the overheated region is moving close to the point. This indicates that the point is located around the boundary of the fusion zone, which encounters a large variation of mechanical properties of the material w.r.t. the high-gradient temperature and thus undergoes a large variation of the shear stress. 

\subsection{Powder-resolved mesoscopic residual stress formation mechanism}

% {
% majorly fig 7
% \begin{itemize}
% \item spatial distribution,
% \item \begin{itemize}
% \item if no fusion zone: stress concentrated around necking;
% \item if fusion zone: inside the fusion zone
% \end{itemize}
% \end{itemize}

% {
% majorly fig 8
% \begin{itemize}
% \item \begin{itemize}
% \item stress: around layer-layer boundary concentrated, not fully released during new layer's fusion
% \item stress's accumulation
% \item origin of the inter-layer force
% \item strain: around fusion zone bottom (strain reseted in the fusion zone)
% \end{itemize}
% \end{itemize}

% majorly fig 7
% \begin{itemize}
% \item spatial distribution,
% \item \begin{itemize}
% \item if no fusion zone: stress concentrated around necking;
% \item if fusion zone: inside the fusion zone
% \end{itemize}
% \end{itemize}}

\begin{figure}[!h]
	\centering
 	\includegraphics[width=18cm]{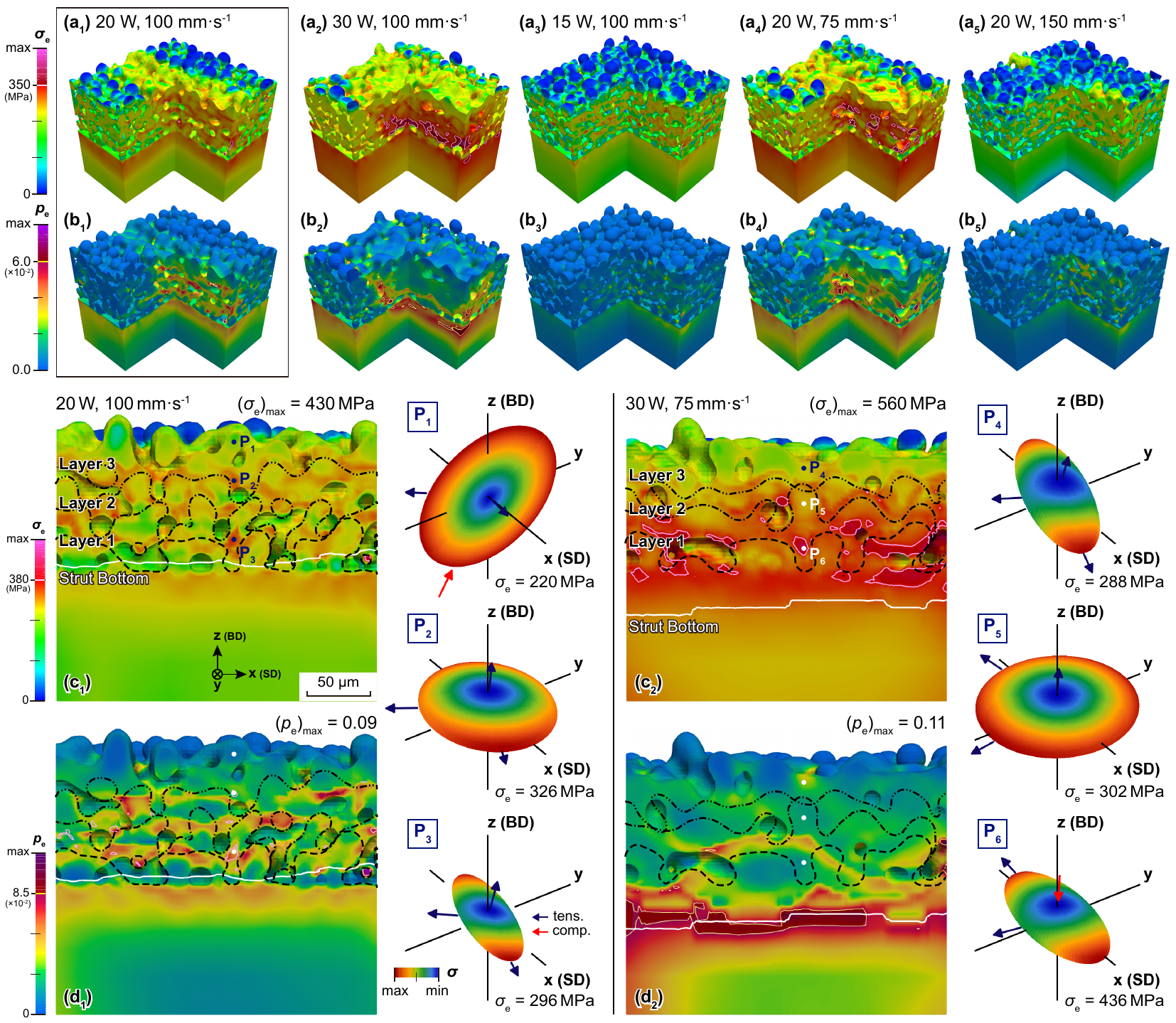}
	\caption{\textbf{Residual stress and effective strain profiles with varying processing parameters}. Simulated profiles of (\ce{a1})-(\ce{a5}) von Mises stress $\vm$ and (\ce{b1})-(\ce{b5}) effective plastic strain $\pe$ in the SLS-processed four-layer powder bed with varying beam power $\Po$ and scan speed $\v$. Sectional profiles of (\ce{c1})-(\ce{c2}) residual von Mises stress $\vm$ and (\ce{d1})-(\ce{d2}) accumulated plastic strain $\pe$ among SLS-processed four-layer powder bed with (\ce{c1})-(\ce{d1}) $\Po=20~\si{W}$ and $\v=100~\si{mm~s^{-1}}$ and (\ce{c2})-(\ce{d2}) $\Po=30~\si{W}$ and $\v=75~\si{mm~s^{-1}}$, respectively. Former surfaces of each layer are denoted by discontinuous black lines, and fused strut bottom is denoted with a solid white line. The Lamé's stress ellipsoids, representing the stress state at selected points P$_1$-P$_6$, are also illustrated with the principle stresses denoted and colored (marine: tension; red: compression).}
	\label{fig:vm}
\end{figure}

For SLS-processed porous microstructures, the residual stress is generally related to stress concentration under thermal loading in the microstructure. In \figref{fig:vm}\ce{a3}, and \ref{fig:vm}\ce{a5}, the particles undergo a low degree of fusion (due to low level of overheating) due to relatively low beam power or high scan speed. The stress concentration mostly occurs at the necking region, leading to accumulated plastic strain and residual stress. In \figref{fig:vm}\ce{a1} and \ref{fig:vm}\ce{a4}, with a higher degree of fusion due to a relatively high beam power or low scan speed, accumulated plastic strain and residual stress tend to concentrate on both necking region and inter-layer region (as indicated by the black lines in \figref{fig:vm}c and \ref{fig:vm}d). Note that the inter-layer boundary has a high surface roughness which also causes stress concentrations. In \figref{fig:vm}\ce{a2}, with the combination of highest $\Po$ and lowest $\v$ within the processing window, residual stress can be observed in the whole fusion zone, which makes it difficult to identify the residual stress concentration. Besides, accumulated plastic strain and residual stress can be observed on the boundary of the substrate and the \nth{1} layer, especially at high beam power. This is because a high beam power generates a much larger fusion zone, which even penetrates into the substrate. When the substrate with continuum material is melted, it generates a large plastic deformation and thus the residual stress, which can be significantly larger than those in the porous microstructure, as shown in \figref{fig:vm}\ce{a2}. These results show that the residual stress is directly related to the degree of fusion and thus related to SLS processing parameters. 

To gain a better understanding of the distribution of residual stress, we analyzed the stress state of a few representative positions in the microstructure, as shown in \figref{fig:vm}c and \ref{fig:vm}d. The former surfaces of fused layers are marked by black lines, and the bottom of the fused strut, a.k.a., the fusion zone boundary (FZB) of the \nth{1} layer, is indicated by white lines. \figref{fig:vm}\ce{c1} and \ref{fig:vm}\ce{d1} show that the residual stress and the accumulated plastic strain are solely concentrated in the necking region and inter-layer region for the case with a low degree of fusion. In \figref{fig:vm}\ce{c2}, with the high degree of fusion (the smallest porosity acheived in the selected processing window), the residual stress and the accumulated plastic strain still tend to be solely concentrated in the inter-layer region, yet less distinguishable compared to the former case. By comparing Lamé’s stress ellipsoids at six points on the boundaries of different layers, it is worth noting that the stress states close to the boundary between layers are very similar, e.g. the stress states of P$_2$ and P$_3$ are similar to that of P$_5$ and P$_6$, no matter how high the degree of fusion is. It means that the residual stress is still formulated due to stress concentration at the inter-layer region. On the other hand, the stress states of P$_1$ and P$_4$ at the top layer are quite different. This is because the residual stress of the top layer is directly determined by the thermal loading which depends on the surface morphology and morphology-induced thermal inhomogeneity.

\begin{figure}[!h]
	\centering
 	\includegraphics[width=16cm]{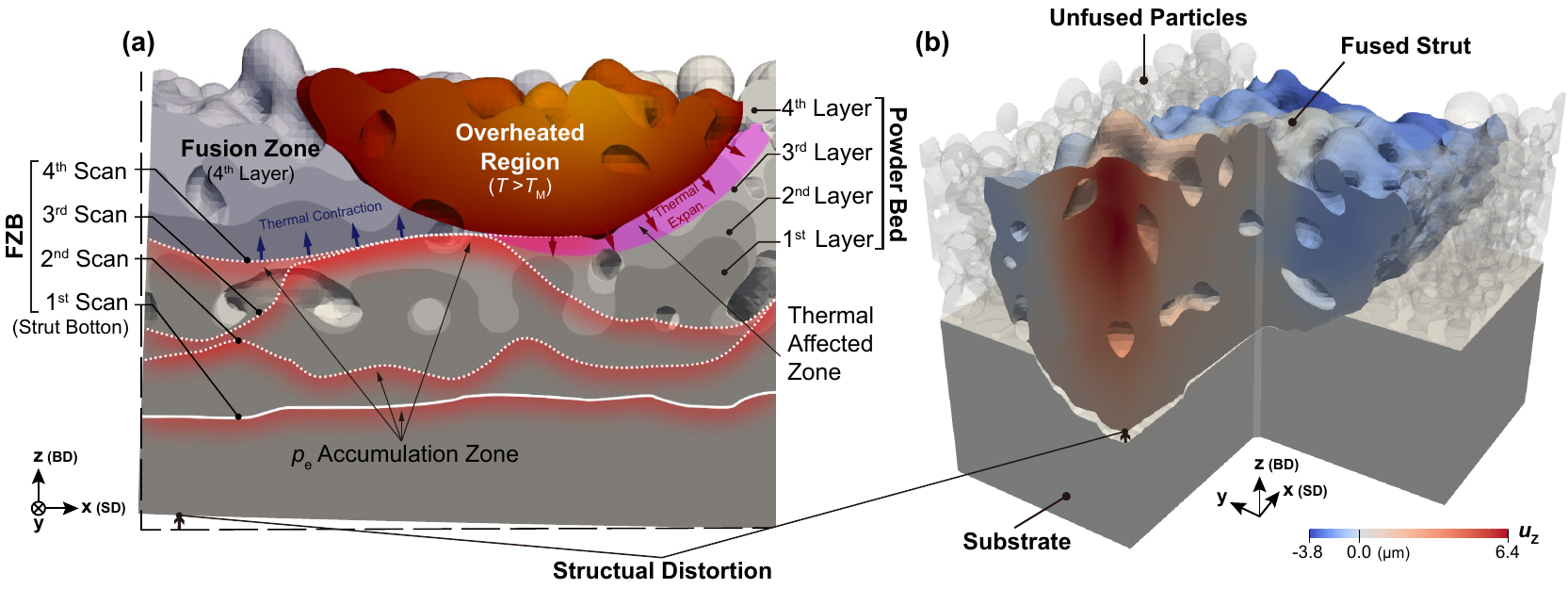}
	\caption{\textbf{Powder-resolved mesoscopic residual stress formation mechanism}. (a) Schematic of the mechanism of powder-resolved mesoscopic residual stress formation, and (b) distorted strut captured in the simulation, colormapped with the displacement along $z$-direction. These schematics are based on the simulated microstructure with $\Po=30~\si{W}$ and $\v = 100~\si{mm~s^{-1}}$. }
	\label{fig:tgm}
\end{figure}

Based on the above observations, we propose a powder-resolved mesoscopic residual stress formation mechanism, which is summarized in the schematic illustrated in \figref{fig:tgm}. The residual stress formulated in porous microstructures manufactured by a multilayer SLS process contains two primary contributions: 
\begin{enumerate}[(i)]
\item Residual stress is directly caused by the stress concentration at the necking region of partially melted particles under thermal loading. The partially melted particles are inter-connected via necking regions, which is the weak link of the microstructure. Therefore, both the thermal expansion in the overheated region and the thermal contraction during the cooling stage cause severe plastic deformation at the necking region, which is one primary source of the accumulated plastic strain and residual stress.
\item Residual stress due to interaction between the upper- and lower-/fused layers in the layerwisely build-up process. In the cooling stage, the shrinkage of the upper-layer after overheating results in tensile stress on itself and compressive stress on the lower-/fused layer, which causes plastic deformation of the porous structure, especially at the inter-layer region with stress concentration due to a high surface roughness, which is the other primary source of the accumulated plastic strain and residual stress, as indicated by the white lines in \figref{fig:tgm}a.
\end{enumerate}

The proposed mechanism is based on the detailed simulation results using the powder-resolved thermo-mechanical model, which evidently demonstrates the formation and distribution of residual stress in the porous structure. The accumulated plastic strain results in structural distortion of the fused strut, which is schematically illustrated in \figref{fig:tgm}a and demonstrated by the contour plot of deformation component $u_z$ in \figref{fig:tgm}b. Compared to the aforementioned TGM and CDS models proposed by Mercelis and Kruth~\cite{mercelis2006residual} correspondingly, which ignored the difference between parts manufactured by SLM and SLS, the partial melting as the main fusion mechanism plays an important role in the residual stress in the microstructure. There are some similarities, the concept of the TGM model also applies to the proposed mechanism because the temperature gradient indeed directly leads to residual stress. However, in the SLS process, since the power bed is fully relaxed before fusion, particles provide very weak constrain on the fusion zone and thus have limited plastic deformation. The plastic deformation is majorly accumulated during the cooling stage due to thermal contraction and stress concentration at the necking region. For the CDS model, the proposed model also considered the shrinkage of the top layer resulting in a tensile residual stress in the top layer and a compressive residual stress in the previous layers. However, the top layer in the simulated SLS process has very complex boundaries with previous layers. These boundaries are the source of the stress concentration which leads to the dominant residual stress at the inter-layer regions. 

Based on the TGM model and the CDS model, Mercelis and Kruth~\cite{mercelis2006residual} also introduced a theoretical model to predict the relationship between residual stress and the part height. Similarly, we can use the proposed powder-resolved model and mesoscopic residual stress formation mechanism to predict the dependence of the residual stress on the microstructure porosity, which is also a characteristic geometric parameter of a porous microstructure. Since the porous structure induces complex inhomogeneity of both material properties and thermal and mechanical loading, and we also need to consider the porosity is directly related to SLS processing parameters, it is more straightforward to propose phenomenological models based on our simulation results to predict the relationship between the residual stress and SLS processing parameters, as will be discussed in the next section.

\section{Discussion}

\subsection{Phenomenological relation for porosity control}\label{sec:pr-poro}

\begin{figure}[!h]
	\centering
 	\includegraphics[width=18cm]{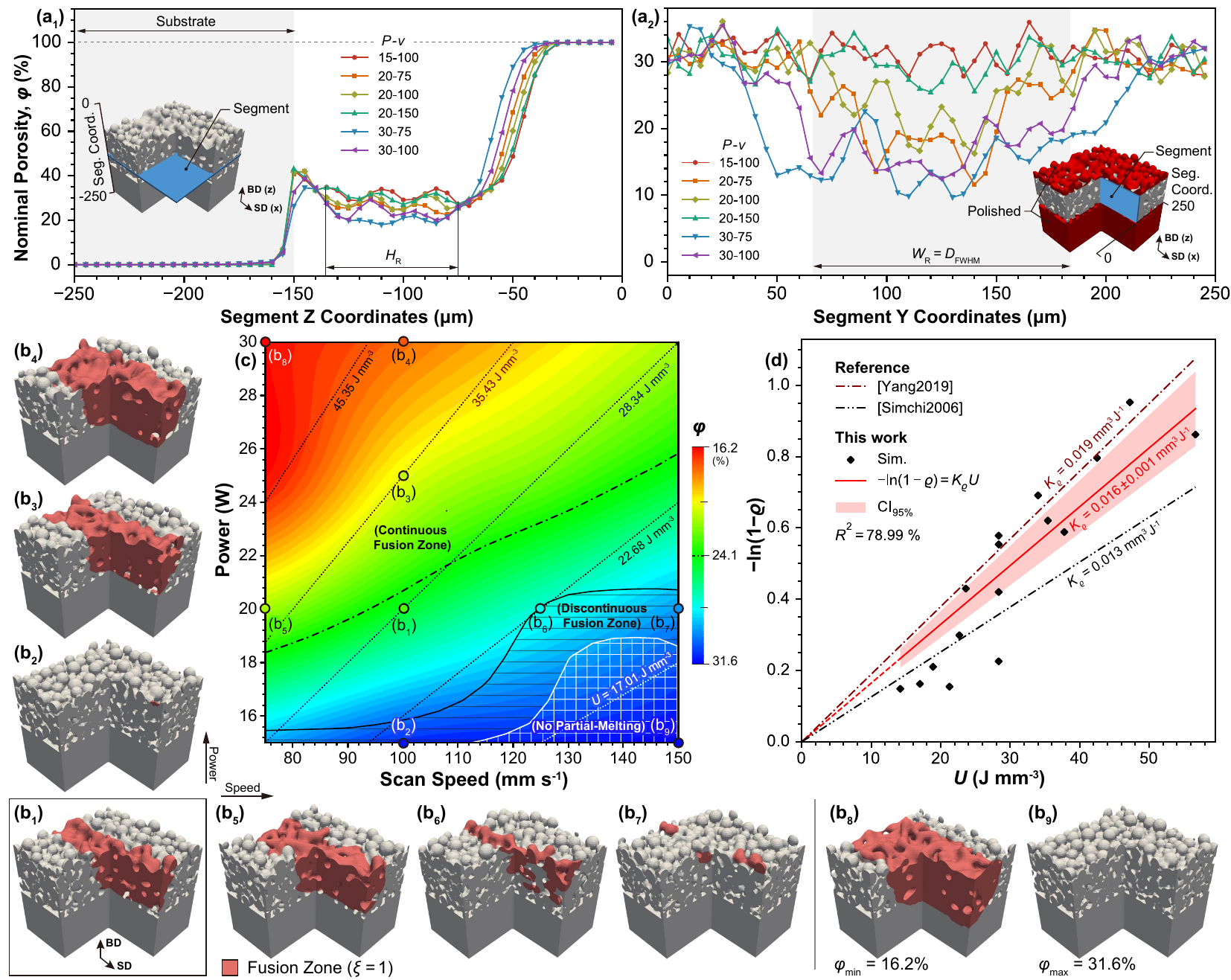}
	\caption{\textbf{Dependence of microstructure on processing parameters}. (a) Segment-wise porosity of the sample produced under various processing parameters along (\ce{a1}) building direction (BD) and (\ce{a2}) perpendicular to scan direction (SD), where the range of the substrate as well as the beam spot size ($D_\mathrm{FWHM}$) are denoted. Representative height ($H_\mathrm{R}$) and width ($W_\mathrm{R}$) for porosity calculation are also selected. (\ce{b1})-(\ce{b9}) SLS-processed microstructures (with fusion zone indicated in red) under varying beam power and scan speed, which are marked as points in the porosity $\Pv$ map (c). The dotted lines represent the volumetric energy input $\uv$ isolines and the dash-dotted line represents the median porosity isoline ($24.1\%$). (d) Phenomenological relation between densification factor $\varrho$ (calculated from porosity) and $\uv$.}
	\label{fig:poro}
\end{figure}

Controlling the porosity of processed sample through the tuning of the processing parameters plays a central role in tailoring the end-up performance of the porous materials in AM, as many properties are determined by (or related to) porosity, including but not limited to mechanical strength, permeability, acoustic/optical absorptivity and various effective conductivities. We start the discussion with phenomenologically relating the porosity to the processing parameters (in this work $\Po$ and $\v$). In this work, a nominal porosity is defined as
\begin{equation}
 \varphi=1-\frac{\int_{\Omega''}\rho\diff \Omega}{{\Omega''}}
 \label{eq:poro}
\end{equation}
with the substance OP $\rho$. $\Omega''$ represents the volume of a post-processed simulation domain with surface and unfused powders sufficiently removed (termed as ``virtual polishing''). \figref{fig:poro}a presents the local porosity that evaluated segment-wise along $z$- (BD) and $y$-direction (perpendicular to SD) to help identify the representative domain with a width $W_\mathrm{R}$ and a height $H_\mathrm{R}$ for porosity calculation, while the complete length ($L_\mathrm{R}=250~\si{\micro m}$) along $x$-direction (SD) is included due to exhibited relatively minor variations in local porosity \cite{zhou20213d}. We then conduct the virtual polishing on simulated multilayer SLS-processed microstructures with $H_\mathrm{R}=60~\si{\micro m}$ (containing the coordinate range $z\in[-135, -75]~\si{\micro m}$) and $W_\mathrm{R}=D_\mathrm{FWHM}$, and proceed calculating their porosity using \eqref{eq:poro} with the post-processed domain size $\Omega''=L_\mathrm{R}H_\mathrm{R}W_\mathrm{R}$. The resulting $\Pv$ map of porosity is in shown \figref{fig:poro}c. Selected microstructures illustrated in Figs. \ref{fig:poro}\ce{b2}-\ref{fig:poro}\ce{b4} for the cases varying $\Po$ with constant $\v$; Figs. \ref{fig:poro}\ce{b5}-\ref{fig:poro}\ce{b7} for the cases varying $\v$ with constant $\Po$. \figref{fig:poro}\ce{b1} is the microstructure under a reference processing parameter ($\Po=20~\si{W}$, $\v=100~\si{mm~s^{-1}}$), while Figs. \ref{fig:poro}\ce{b8} and \ref{fig:poro}\ce{b9} are respectively the one with minimum and maximum porosity. It demonstrates that improved densification can be achieved by either increasing $\Po$ or decreasing $\v$, where smoothed surface morphology implies an enhanced partial melting. Porosity drops from 31\% to 18\% for the increase of $\Po$ from $15$ to $30~\si{W}$ ($\v=100~\si{mm~s^{-1}}$) and from 29\% to 21\% for the decrease of $\v$ from $150$ to $75~\si{mm~s^{-1}}$ ($\Po=20~\si{W}$). The presented tendencies imply a possible allometric relation $\varphi \propto P^{-m} v^n$ with positive indices $m$ and $n$. On the other hand, combining beam power and scan speed as one characteristic quantity, $P/v$ is widely adapted to define a specific energy density, notably the volumetric energy input as
\begin{equation}
 \uv = \frac{\Po}{H W \v},
\end{equation}
where $H$ is the powder layer thickness (here $H=60~\si{\micro m}$) and $W$ is the scan track width (here $W$ takes $D_\mathrm{FWHM}$). Evidently, $\varphi$ exhibits an overall decrease tendency with rising $\uv$, meaning lower porosity can be achieved as higher specific energy input. 

We then examined the simulation results with a proposed phenomenological relation following Refs. \cite{simchi2006direct,yang2019npj}, read as
\begin{equation}
 \ln[1-\varrho(\varphi)]=- K_\varrho \uv
\end{equation}
with a densification factor defined as
\begin{equation*}
 \quad \varrho(\varphi) \equiv \frac{\varphi_{0} - \varphi}{\varphi_{0} - \varphi_\text{min}}.
\end{equation*}
$\varrho$ indicates the ratio between a reduced porosity w.r.t. the maximum porosity reduction in the chosen processing window, as $\varphi_0$ and $\varphi_\text{min}$ represent the initial and minimum achieved porosity, respectively. $\varphi_\text{min}$ usually varies between 3\% to 30\% for metallic materials \cite{simchi2006direct}. In pursuit of uniformity, we have chosen $\varphi_\text{min}=3\%$ in this study, aligning with our previous single-layer SLS simulation \cite{yang2019npj}. $\varphi_0$, however, is difficult to be determined for multilayer SLS simulation since there is sorely one or two particles in thickness for every new layer while the old layers have been fused. In this sense, we evaluate $\varphi_{0}$ by three routes:
\begin{enumerate}[(i)]
\item Assuming that the porosity of the powder bed region, which is away from the beam spot (i.e., without significant densification), is the initial one of the powder bed, $\varphi_{0}$ can be read as the converged value from segment-wise porosity evaluation along $y$-direction (\figref{fig:poro}\ce{a2}), which is $\varphi_{0}=27.8\%\sim33.6\%$. 

\item As the particle size distribution in this work is assumed to be Gaussian-type, $\varphi_{0}$ is then estimated by statistic random-close-packing model of spherical particle as $\varphi_0 = 0.366 - 0.0257\left({\varsigma_d}/{\mu_d}\right) $ with $\mu_d$ and $\varsigma_d$ the mean and standard deviation of the particle diameter \cite{desmond2014influence, yang2022validated}. Taking $\mu_d=20~\si{\micro\meter}$ and $\varsigma_d=5~\si{\micro\meter}$, it can be calculated as $\varphi_0 = 36.0 \%$.

\item We also piled multiple densification-free powder stacks using DEM method with the same domain volume fraction of the particles ($48.0\%$) as the overall one of the multilayer SLS simulation, in which each layer deposits powders of 12\% domain volume fraction. The measured nominal porosity is thereby regarded as the initial one, as $\varphi_0 = 40.0\%\sim40.5\%$. 
\end{enumerate}
$\varphi_0$ from route (i) directly reflects the porosity of on-site unfused powders, but the influences from the potential necked particles due to thermal processes cannot be well eliminated. $\varphi_0$ from route (ii) is based on the statistics of random-closed-packed particles, which is rather ideal compared to the practical powder bed created by powder spreading. Since route (iii) creates particle stacks via simulating the deposition in powder spreading, the calculated $\varphi_0$ may be still inflated as the morphological variability of the deposition surface is missing. Considering all of these factors, we have selected $\varphi_0 = 36.0\%$, which stands at the midpoint among all the assessed values.

In \figref{fig:poro}d, linear regression between $-\ln(1-\varrho)$, calculated from simulated $\varphi$, and $\uv$ is presented. For comparison, regressions on data by single-layer SLS simulation and experiment are also illustrated \cite{simchi2006direct, yang2019npj}. The regressed line together with the 95\% confidence interval ($\cib$) of the multilayer SLS simulations is right in between those of the single-layer SLS simulation and experiments. As the coefficient $K_\varrho$ is related to the material and the size distribution of the powders, the multilayer result $K_\varrho=0.016\pm0.001~\si{mm^3~J^{-1}}$ demonstrates an improved coherence with the experimental one $K_\varrho=0.013~\si{mm^3~J^{-1}}$ compared the single-layer one $K_\varrho=0.019~\si{mm^3~J^{-1}}$, which can suffer from inflated porosity mostly due to insufficient volume microstructure for porosity calculation. {It is also worth noting that this relation between $-\ln(1-\varrho)$ and $\uv$ is examined on the microstructures both with and without fusion zone. To characterize the formation of the fusion zone, a historical indicator $\xi$ was added in the system to emulate the phenomenological fusion of the strut during the multilayer SLS process following our former work \cite{yang2022validated, yang2023npj}, which is initialized as zero and irreversibly transitions as one once $T\geq\Tm$. In \subssfigref{fig:poro}{b1}{b9} the fusion zones are denoted for selected $\Po$ and $v$. The map can be thereby divided into characteristic regions where a continuous/discontinuous fusion zone or no fusion zone is formed. Increasing $U$ from 17.01 to 22.68 \si{J~mm^{-3}}, the transition from microstructures without fusion zone to with continuous fusion zone also implies the switch of densification
mechanism from pure solid-state sintering to partial-melting--induced fusion (similar to the liquid-state sintering). This may explain the change of the tendency of $-\ln(1-\varrho)$ when $\uv<22.68~\si{J~mm^{-3}}$ in \subfigref{fig:poro}{d}.}

 % However, as the the $\varphi$ contours intersect with $\uv$ isolines, the porosity seems not uniquely dependent on $\uv$
 % Following the discussion above, we take the $h_\mathrm{pb} = 60 \si{\micro\meter}$ as the thickness $h$, and the $D_\mathrm{FWHM}=117.6~\si{\micro\meter}$ as the width $w$. As for calculation of $\varrho$, $\varphi_\text{min}$ is the minimum attainable porosity in a sintered part, 

%To represent our findings from the selected processing parameters in a wider range, we choose $\varphi_\text{min} = 15.0\%$, which is close to the simulated minimum porosity $\varphi_\mathrm=16.9\%$ and still lies in the range suggested by Ref. \cite{simchi2006direct}.

% \begin{figure}[!h]
% 	\centering
% 	\includegraphics[width=10cm]{figures/fig_Tprobe.pdf}
% 	\caption{xxx.}
% 	\label{fig:Tp}
% \end{figure}

\subsection{Phenomenological relations for residual stress and plastic strain control}
\begin{figure}[!t]  
	\centering
 	\includegraphics[width=18cm]{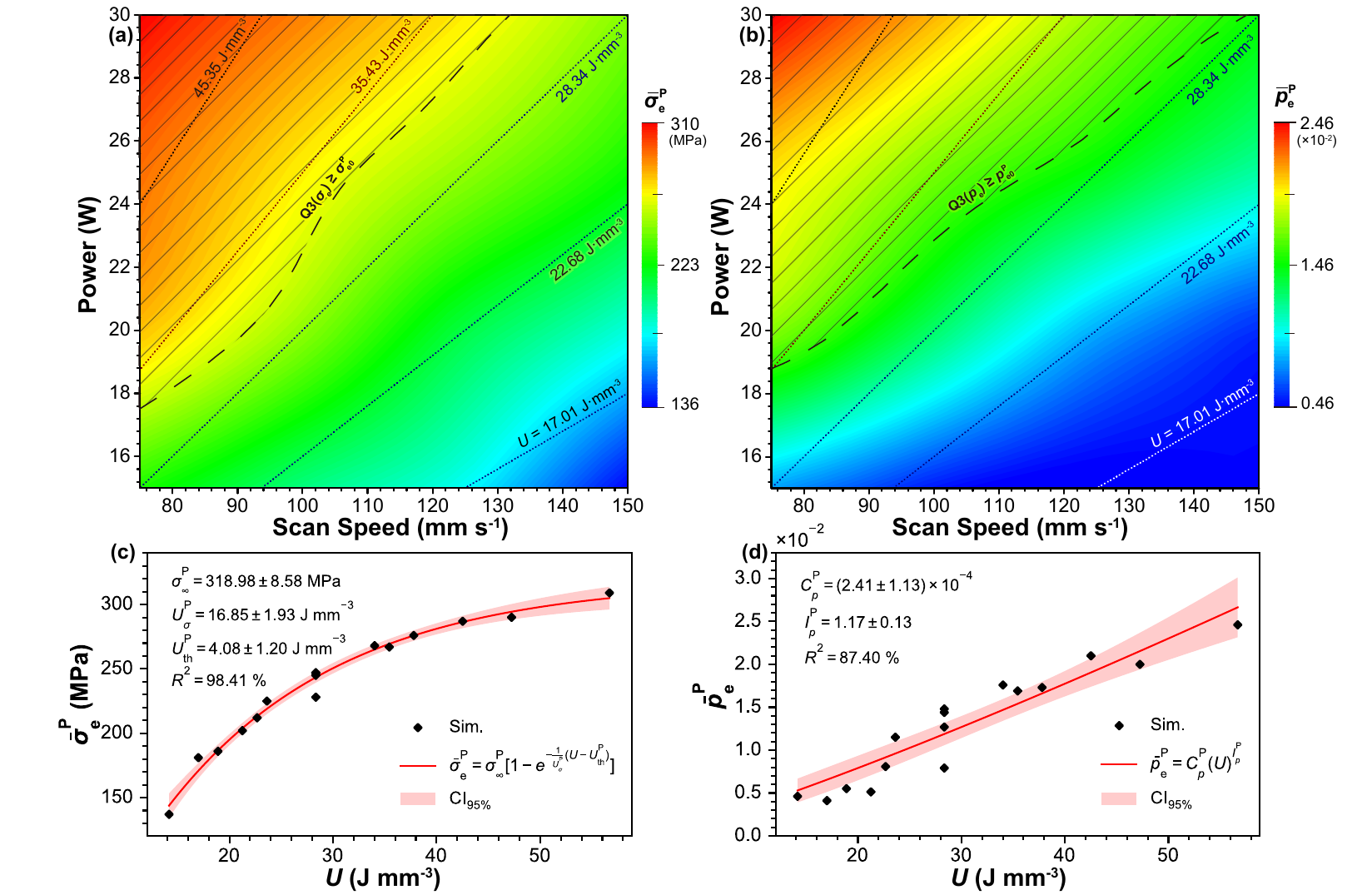}
	\caption{\textbf{Dependence of average residual stress and effective plastic strain microstructure on processing parameters inside the powder bed (PB)}. $\Pv$ maps of (a) PB-averaged residual stress $\barvm^\mathrm{P}$ and (b) PB-averaged plastic strain $\barpe^\mathrm{P}$. The dotted lines maps represent the volumetric energy input $\uv$ isolines. $\Pv$ pairs for simulated profiles in \figref{fig:vm}a and \ref{fig:vm}b are denoted correspondingly. $\Pv$ regions with third quantiles $\QQQ({\vm})$ and $\QQQ({\pe})$ beyond the referencing PB-averages $\sigma^\mathrm{P}_\mathrm{e0}=247~\mathrm{MPa}$ and $p^\mathrm{P}_\mathrm{e0}=0.012$ of the processed microstructure under $\Po=20~\si{W}$ and $\v=100~\si{mm~s^{-1}}$ are also denoted. 
 Nonlinear regressions of (c) $\barvm^\mathrm{P}$ and (d) $\barpe^\mathrm{P}$ on $\uv$ with the regression parameters indicated correspondingly.} 
	\label{fig:rmap}
\end{figure}

Taking the PB-averaged residual von Mises stress $\barvm^\mathrm{P}$ and effective plastic strain $\barpe^\mathrm{P}$ (defined in \eqref{eq:pbavg}) at the end of the simulations, \figref{fig:rmap} presents the distributions of $\pbvm$ and $\pbpe$ w.r.t. the $\Po$ and $\v$ in the chosen processing window. Generally, the rise in $\pbvm$ corresponds to an increase in the specific energy input $\uv$ (i.e., increase $\Po$ with maintained $\v$ or increase $\v$ with maintained $\Po$), resulting in further concentrated residual stress around concave features (surface depressions and particle sintering necks) and across layers, as already presented in \figref{fig:vm}a and \ref{fig:vm}b. 
{Meanwhile, as the locally concentrated residual stresses and plastic strain are evident in the processed powder bed, understanding how these local quantities are distributed and how much effect the extreme values (here the local maximums
) impose on the evaluated averages $\barvm^\mathrm{P}$ and $\barpe^\mathrm{P}$ becomes essential. In that sense, we sampled all nodal points in the processed powder bed and performed statistics on $\vm$ and $\pe$ of these points. The probability distribution of local $\vm$ and $\pe$ are presented in Supplementary Fig. 1a. The third quartile (Q3) is employed to characterize the concentration of the local quantities. Then, $\Pv$ regions with Q3($\vm$) and Q3($\pe$) correspondingly beyond the referencing values, which are PB-averages $\sigma^\mathrm{P}_\mathrm{e0}=247~\mathrm{MPa}$ and $p^\mathrm{P}_\mathrm{e0}=0.012$ from the case $\Po=20~\si{W}$ and $\v=100~\si{mm~s^{-1}}$, are denoted in \figref{fig:rmap}a and \ref{fig:rmap}b (the complete $\Pv$ maps for Q3($\vm$) and Q3($\pe$) are shown in Supplementary Fig. 2). These can be statistically interpreted as that the processed powder bed with $\Po$ and $\v$ selected within the regions have at least 25\% local substances obtaining $\vm$ and $\pe$ beyond the referencing $\sigma^\mathrm{P}_\mathrm{e0}$ and $p^\mathrm{P}_\mathrm{e0}$, respectively, while the rest of local substances have adequate $\vm$ and $\pe$ up to these referencing values. In other words, the SLS processes with $\Po$ and $\v$ selected from these regions tend to have processed powder bed with relatively higher localized residual stresses and accumulated plastic strains.
} 

To understand the relationship between $\pbvm$ and $\uv$, we interpreted $\pbvm$ as the stored mechanical energy density after the multilayer SLS processes, which can be regarded as the residue of the energy density imported via the beam-induced thermal effect in the powder bed after all types of in-process dissipation. In this regard, the nonlinear regression analysis of an energy conversion law $\pbvm=\sigma^\mathrm{P}_\infty \{1-\exp[{{-\frac{1}{U^\mathrm{P}_\sigma}(U-U^\mathrm{P}_\mathrm{th})}}]\}$ was conducted on the simulation data. The parameters $U^\mathrm{P}_\mathrm{th}$ and $\sigma^\mathrm{P}_\infty$ adopt the physical meanings as the volumetric energy input at stress-free state and the saturated stress at infinity energy input, respectively, and $\sigma^\mathrm{P}_\infty/U^\mathrm{P}_\sigma$ is the increasing rate (slope) of $\pbvm$ vs. $\uv$ at stress-free state, as shown in the inset of \figref{fig:rmap}c. The result in \figref{fig:rmap}c presents a high correlation ($R^2=98.41\%$) between simulated $\pbvm$ vs. $\uv$ with narrow confidence interval, demonstrating the applicability of proposed energy conversion law in predicting the residual stress in the powder bed. 
% The regression analysis also delivers a predicted $U^\mathrm{P}_\mathrm{th}=4.08\pm1.20~\si{J~mm^{-3}}$, implying a threshold energy input to obtain a stress-free microstructure; and a $\sigma^\mathrm{P}_\infty=318.98\pm8.58~\si{MPa}$, implying the highest possible residual stress in the simulation condition. These quantities request further examination and validation in future numerical and experimental studies.

Unlike the residual stress, the effective plastic strain $\pe$ is a measure of cumulative plastic deformation at any given moment during the process and lacks a clear physical picture of its relationship with volumetric energy input $\uv$. Therefore, the nonlinear regression analysis of an allometric scaling law $\pbpe=C^\mathrm{P}_p (U)^{I^\mathrm{P}_p}$ with parameters $C^\mathrm{P}_p$ and $I^\mathrm{P}_p$ was conducted to phenomenologically relate $\pbpe$ to $\uv$, as shown in \figref{fig:rmap}d. The analysis gives a relatively low correlation coefficient $R^2=87.4\%$ of the $\pbpe(\uv)$ relation compared with one of $\pbvm(\uv)$, with an expanded confidence interval at high $\uv$ range. Regressed $I^\mathrm{P}_p=1.17\pm0.13$ indicates an almost linear scalability of accumulated plastic strain in the processed powder bed on energy input via beam scan. It is worth noting that the proposed scaling law can be challenged as $\uv$ appears to not being able to uniquely identify $\pbpe$, in other words, $\uv$ of certain value can correspond to multiple $\pe$ value, as also depicted in \figref{fig:rmap}d. Nonetheless, this scaling law can find it feasible in estimating strength of in-process plastification of the microstructure at the given specific energy input.

% volumetric ratios between the region where $\vm>\barvm$ and $\pe>\barpe$ and the whole powder bed are defined to characterize the degree of localized concentration, i.e., $\psi_{\vm>\barvm}=\Omega(\vm>\barvm)/\Omega^\mathrm{P}$ and $\psi_{\pe>\barpe}=\Omega(\pe>\barpe)/\Omega^\mathrm{P}$. } 

\begin{figure}[!t]  
	\centering
 	\includegraphics[width=18cm]{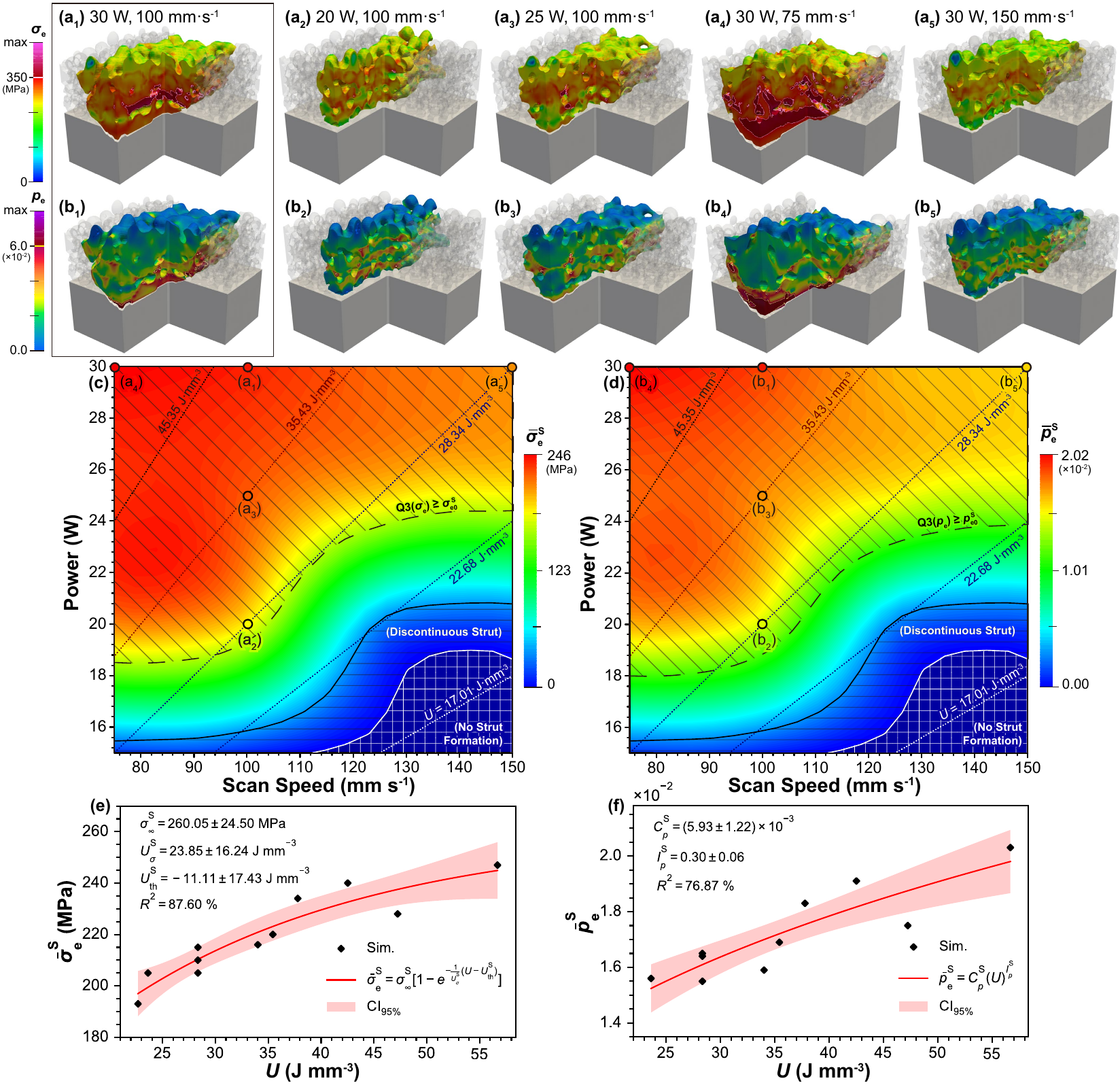}
	\caption{\textbf{Dependence of average residual stress and effective plastic strain microstructure on processing parameters inside the fused strut}. Simulated profiles of (\ce{a1})-(\ce{a5}) residual von Mises stress $\vm$ and (\ce{b1})-(\ce{b5}) effective plastic strain $\pe$ in the distorted fused strut with varying beam power $\Po$ and scan speed $\v$, which are marked as points in the $\Pv$ maps of (c) strut-averaged residual stress $\barvm^\mathrm{S}$ and (d) strut-averaged plastic strain $\barpe^\mathrm{S}$, respectively. The dotted lines maps represent the volumetric energy input $\uv$ isolines. $\Pv$ regions with third quantiles $\QQQ({\vm})$ and $\QQQ({\pe})$ beyond the referencing strut-averages $\sigma^\mathrm{S}_\mathrm{e0}=215~\mathrm{MPa}$ and $p^\mathrm{P}_\mathrm{e0}=0.016$ of the the processed microstructure under $\Po=20~\si{W}$ and $\v=100~\si{mm~s^{-1}}$ are also denoted along with ones indicating different geometries of fused struts. Nonlinear regressions of (e) $\barvm^\mathrm{S}$ and (f) $\barpe^\mathrm{S}$ on $\uv$ with the regression parameters indicated correspondingly.} 
	\label{fig:rmap2}
\end{figure}

Since both $\pbpe$ and $\pbvm$ consider a complete powder bed with unfused particles, we further defined two average quantities that only take the residual stress and effective plastic strain inside the fused strut into account, denoted respectively $\fsvm$ and $\fspe$, which are calculated as 
\begin{equation}
 \fsvm = \frac{\int_{\Omega} \xi \sigma_\mathrm{e} \diff \Omega}{\int_{\Omega} \xi \diff \Omega}, \quad
 \fspe = \frac{\int_{\Omega} \xi p_\mathrm{e} \diff \Omega}{\int_{\Omega} \xi \diff \Omega}
\end{equation}
with the fusion zone indicator $\xi$. \figref{fig:rmap2} presents the distribution of $\fsvm$ and $\fspe$ w.r.t. $\Po$ and $\v$ in the chosen processing window. Contrasting with those shown in \figref{fig:rmap}, a notable distinction in the maps presented in \figref{fig:rmap2}c and \ref{fig:rmap2}d is the emergence of regions where the selected $P$ and $v$ fail to form continuous fused strut. Meanwhile, the profiles of $\fsvm$ and $\fspe$ receive significant influences from the size of the strut. Comparing \figref{fig:rmap2}\ce{a1} with \ref{fig:rmap2}\ce{a2}-\ref{fig:rmap2}\ce{a3} and \ref{fig:rmap2}\ce{a4}-\ref{fig:rmap2}\ce{a5}, the increase of $\fsvm$ follows the direction of increasing $\uv$ as well, which simultaneously leads to the enlargement of the fused strut. As depicted in Figs. \ref{fig:vm}c and \ref{fig:rmap2}a, concentrated residual stress is primarily found within the upper-layers of the strut. When the strut size is enlarged, it encompasses a larger volume with elevated maximum $\vm$ and high-$\vm$ region, resulting in an increased $\fsvm$ with the rise in $\uv$. $\fspe$ exhibits similar tendency w.r.t. $\Po$ and $\v$ as $\fsvm$, with an increase in $\uv$ resulting in an increase in $\fspe$. Nonetheless, regions with highly accumulated $\pe$ locates at the bottom of each layer's fusion zone, as illustrated in Figs. \ref{fig:vm}d and \ref{fig:rmap2}b. At high $\uv$, such accumulation intensifies, especially at the bottom of the strut (which is also the bottom of the initial layer's fusion zone). Simultaneously, enlarged depth of a fusion zone indicates an extended remelting in the fused lower-layers, which removes some accumulated $\pe$ by the process or former layer within the strut, as shown in \figref{fig:tgm}a. It also concentrates the high-$\pe$ region further to the bottom of the strut, notably the profile in \figref{fig:rmap2}\ce{b4}. Eventually, the rise of $\fspe$ w.r.t. $\uv$ is relatively less ``rapid'' than the one of $\fsvm$, evident in slightly sparser contours in \figref{fig:rmap2}d compared to \ref{fig:rmap2}c. {Statistics of the local $\vm$ and $\pe$ in the fused strut are also performed (Supplementary Fig. 1b), and the $\Pv$ regions characterizing Q3($\vm$) and Q3($\pe$) beyond the referencing values, which are strut averages $\sigma^\mathrm{S}_\mathrm{e0}=215~\mathrm{MPa}$ and $p^\mathrm{S}_\mathrm{e0}=0.016$ from the same case $\Po=20~\si{W}$ and $\v=100~\si{mm~s^{-1}}$, are denoted in the \figref{fig:rmap2}c and \ref{fig:rmap2}d (the complete $\Pv$ maps for Q3($\vm$) and Q3($\pe$) are shown in Supplementary Fig. 3). As many local points outside the fusion zone are excluded from the statistics, extended regions of $\QQQ(\vm)>\sigma^\mathrm{S}_\mathrm{e0}$ and $\QQQ(\pe)>p^\mathrm{S}_\mathrm{e0}$ are evident, demonstrating that more $\Pv$ combinations can lead to the fused strut containing at least 25\% local $\vm$ and $\pe$ are beyond the averages $\sigma^\mathrm{S}_\mathrm{e0}$ and $p^\mathrm{S}_\mathrm{e0}$ of the same chosen reference case. One can also conclude that most formed continuous struts more or less have localized residual stresses and accumulated plastic strains, based on the observation of narrow un-marked $\Pv$ regions between one with the discontinuous strut and one with $\QQQ$ of local quantities beyond the chosen references. In this sense, secondary thermal treatment should be positively considered to anneal the residual stresses inside the SLS-processed struts.} 

Nonlinear regression analyses were also conducted on $\fsvm$ and $\fspe$ employing the proposed energy conversion law for residual stress and scaling law for effective plastic strain. Results are correspondingly presented in \figref{fig:rmap2}e and \ref{fig:rmap2}f. Notably, comparing to those of $\pbvm(\uv)$ and $\pbpe(\uv)$, the correlation of $\fsvm(\uv)$ and $\fspe(\uv)$ present decline, respectively, with enlarged confidence interval at both low and high $\uv$ ranges. This attributes to the removal of the influences from the unfused particles. Moreover, for $\fsvm(\uv)$, regressed threshold energy input presents a negative value as $U_\mathrm{th}^\mathrm{S}=-11.11\pm17.43~\si{J~mm^{-3}}$, indicating a required energy output to achieve stress-free state. This is explainable as the $\vm$ already exists in a just-formed strut. In other word, primary SLS process (i.e., the SLS without subsequent thermal post-process to release residual stress) cannot achieve samples with stress-free strut. Regressed saturated stress $\sigma^\mathrm{S}_\infty=260.05\pm24.5~\si{MPa}$ also presents decline comparing to the regressed one on PB-averaged data. For $\fspe(\uv)$, a reduced regressed innndex of the scaling law $\fspe(\uv)$, i.e., $I^\mathrm{S}_p=0.30\pm0.06$ in \figref{fig:rmap2}f, demonstrates a sublinear scalability of $\fspe(\uv)$ compared with one of $\pbpe(\uv)$, which is almost linear. It also reflects a reduced growth rate of $\fspe$ at high $\uv$, and the underlying intensified remelting, which removes some accumulated $\pe$ within the strut while further concentrates $\pe$ around the strut bottom, and the comparably faster increase in strut size shall be the reason. Nonetheless, information conveyed by $\fsvm(\uv)$ and $\fspe(\uv)$ is more relevant for practical application, as unfused particles shall be removed during post-process of a practical SLS. Further examination and validation of the proposed laws for residual stress and plastic strain w.r.t. specific energy input are expected in the future numerical and experimental studies. 

 \subsection{Conclusion}
In this work, we proposed a powder-resolved multilayer simulation scheme for producing porous materials using selective laser sintering, combing FEM-based non-isothermal phase-field simulation and thermo-elasto-plastic simulation with temperature-dependent material properties. This work has presented the mesoscopic evolution of stress and plastic strain on a transient thermo-microstructure under various beam power ($\Po$) and scan speed ($\v$). Process-property relationships between porosity, residual stress and effective plastic strain and the volumetric energy input ($\uv\propto \Po/\v$) have also been demonstrated and discussed. 
The following conclusions can be drawn from the present work:
\begin{enumerate}[(i)]
 \item We proposed in this work a new powder-resolved mesoscopic residual stress formation mechanism for porous materials manufactured by the SLS process, which lead to the structural distortion appeared in fused strut. It was demonstrated with simulation results that the stress concentration at the necking region of the partially melted particles and inter-layer region between different layers provide dominant accumulated plastic strain and residual stress in the porous material.
 \item Based on the proposed residual stress formation mechanism, we examined the proposed phenomenological relation between the porosity (densification) and the volumetric energy input $\uv$. Regression analysis 
 on the resulting porosity from multilayer SLS simulations suggested an improved coherence with the experimental data, as the regressed densification coefficient $K_\varrho=0.016\pm0.001~\si{mm^3~J^{-1}}$ in this work is compared to the experimental $K_\varrho=0.013~\si{mm^3~J^{-1}}$ and the one from our former single-layer simulation $K_\varrho=0.019~\si{mm^3~J^{-1}}$. 
 \item Two types of average quantities, namely PB-averaged ($\pbvm$ and $\pbpe$) and strut averaged ones ($\fsvm$ and $\fspe$), were defined to characterize the residual stress and plastic strain within the powder bed and fused strut, correspondingly. The relationships between these quantities and volumetric energy input ($\uv$) are unveiled by conducting nonlinear regression analyses. The average residual stress ($\pbvm$ and $\fsvm$) relates to $\uv$ by the energy conversion law, while the average effective plastic strain ($\pbpe$ and $\fspe$) relates to $\uv$ by the allometric scaling law. 
 \item Attributing to the removal of influences from unfused particles, the correlation of the relations $\fsvm(\uv)$ and $\fspe(\uv)$ present drops compared with one of $\pbvm(\uv)$ and $\pbpe(\uv)$, respectively. Saturation behavior is observed on both $\pbvm(\uv)$ and $\fsvm(\uv)$, while the linear scalability in $\pbpe(\uv)$ degenerates into sublinear one in $\fsvm(\uv)$, demonstrating a reduced growth rate of $\fspe$ at high $\uv$. 
\end{enumerate}
Despite the feasibility of the multilayer simulation scheme in recapitulating mesoscopic formation of porosity, residual stress and plastic strain under given processing parameters; and the proposed mechanism in explaining the structural distortion of SLS-produced samples, several points should be further examined and discussed in future works: 
% Despite the present findings, several points should be further examined and discussed in future works:
\begin{enumerate}[(i)]
 \item The present work omits the consideration of chronological-spatial distribution of the thermal-elasto-plastic properties among polycrystals, as the properties such as elasticity and crystal plasticity vary from grain to grain with distinct orientations.
 \item The present findings are examined at relatively low specific energy input $\uv$ and, correspondingly, low generated residual stress and accumulated plastic strain. Pore formation is also limited to lack-of-fusion mechanism. It is anticipated to conduct further simulations with relatively high energy input, where the mechanisms like keyholing co-exist with high residual stress and plastic strain. Extension of the proposed mechanism into the high-$\uv$ range together with the following examination and validation are also expected. 
 % \item The present findings are only examined at relatively low specific energy input and, correspondingly, low generated residual stress, as the SLS is chosen in this work. It is anticipated to conduct the simulations with relatively high energy input, like SLM, and examine the influences of magneto-elastic coupling on the magnetic hysteresis with comparably higher residual stress cases. Influences on residual stress development and, eventually, the coercivity of manufactured permalloy from multilayer and multitrack AM strategies should also be examined.
\end{enumerate}
% integrated the mesoscopic coupled thermo-structural evolution and associated thermo-elasto-plastic behaviors, nanoscopic $\gpgpd$ transition, and associated Ni segregation and micromagnetics with magneto-plastic coupling into one scheme to demonstrate the using SLS under a scenario that is close to the practical experiments. 

\section{Methods}

% This is based on the fact that the propagation of elastic wave is generally faster than the mechanisms such as diffusion and 
% {This is based on the fact that the propagation of elastic wave
% kinetic parameters of microstructure evolution (e.g., diffusivity and grain boundary mobility) are exponentially local temperature profile exponentially alter the magnitude of diffusivity and grain boundary mobility, leading to vast differences in the progress of microstructure evolution.} 

\subsection{Non-isothermal phase-field model for multilayer SLS processes}\label{sec:npt}

The non-isothermal phase-field model employed in this work to simulate the multilayer is based on our former works \cite{yang2019npj, yang2020investigation, zhou20213d, yang2023npj}. For the sake of completeness, in this section we summarize the essentials of the employed model. {For clarity, vectors are denoted by italic symbols with accented arrow (e.g., $\vec{v}$). Bold symbols (e.g., $\mathbf{K}$, $\mathbf{M}$, $\stress$, $\strain$) are for \nth{2}-order tensors, and blackboard bold symbols (e.g., $\Cten$) are for \nth{4}-order ones.}

% \begin{figure}[t]
% \centerline{\includegraphics[width=0.8\textwidth]{figures/Fig_microstructure_PFS.png}}
% \caption{multiwell}
% \label{mw1}
% \end{figure}

% The free energy functional $\mathscr{F}$ of the simulation domain $\Omega$, depending on the local temperature $T$ and OPs, is the formulated as as
% 	\begin{equation}
% 	\begin{split}
% 	\mathscr{F}(T,\rho,\{\eta_i\})=\int_\Omega\left[f(T,\rho,\{\eta_i\}) + 
% 	\frac{1}{2} T \kappa_\rho \left| \nabla \rho \right|^2 +
% 	\frac{1}{2} T \kappa_\eta \sum_i \left| \nabla \eta_i \right|^2\right] \text{d}\Omega,
% 	\label{eqi2}
% 	\end{split}
% 	\end{equation}
% where the 

The model adopts both conserved and non-conserved OPs for representing the microstructure evolution of polycrystalline material. The conserved OP $\rho$ indicates the substance, including the unfused and partially melted regions, while the non-conserved OPs $\{\eta_i\},~i=1,2,\dots$ distinguish particles with different crystallographic orientations. To guarantee that the orientation fields ($\{\eta_i\}$) get valued only in the substance ($\rho=1$), a numerical constraint $\sum_i \eta_i + (1-\rho) = 1$ is imposed to the simulation domain. This constraint also applies to the substrate where $\rho=1$ always holds.

The thermo-structural evolution is governed by
\begin{empheq}[left=\empheqlbrace]{align}
&\frac{\partial\rho}{\partial t}=
\nabla \cdot \left\{ \mathbf{M}\cdot \nabla \left[\frac{\partial f}{\partial\rho}-\underline{\kappa}_\rho(T)\nabla^2\rho \right]\right\}
,\label{eqk1}\\
&\frac{\partial\eta_i}{\partial t}=-L\left[\frac{\partial f}{\partial\eta_i}-\underline{\kappa}_\eta(T) \nabla^2\eta_i\right]
,\label{eqk3}\\
&c_{\mathrm{r}}\left[ \frac{\partial T}{\partial t} -\vvec\cdot\nabla T\right] =\nabla \cdot\left[ \mathbf{K} \cdot \nabla T\right] +q,\label{eqk5}
\end{empheq}
where the local free energy density is formulated as
\begin{equation}
\begin{split}
f(T,\rho,\{\eta_i\})&= h_\subst(\rho) f_\text{ht}(T)+
\underline{W}_\sf(T)\left[\rho^2(1-\rho)^2 \right] + \\
&\underline{W}_\gb(T)\left[\rho^2+6(1-\rho)\sum_i\eta_i^2 - 4(2-\rho)\sum_i\eta_i^3 + 3\left(\sum_i\eta_i^2 \right)^2 \right],\\
f_{\mathrm{ht}}(T)& =c_{\mathrm{r}}\left[\left(T-T_\mathrm{M}\right)-T \ln \frac{T}{T_\mathrm{M}}\right] +f_\mathrm{ref}^{T_\mathrm{M}} -\frac{T-T_{\mathrm{M}}}{T_{\mathrm{M}}} \mathcal{L}_\mathrm{M}h_{\mathrm{M}}(T). 
\end{split}
\label{eqm2}
\end{equation}
% and the temperature-dependent coefficients of Landau polynomial and gradient terms are
% \begin{equation}
% \begin{split}
% \underline{W}_\sf(T)=W_\sf\tau_\sf(T),\quad\underline{W}_\gb(T)=W_\gb\tau_\gb(T),\\
% \underline{\kappa}_\sf(T)=\kappa_\sf\tau_\sf(T),\quad\underline{\kappa}_\gb(T)=\kappa_\gb\tau_\gb(T).
% \end{split}
% \end{equation}
Here $f(T,\rho,\{\eta_i\})$ contains multiple local minima, representing the thermodynamic stability of the states, i.e., substance, atmosphere/pore and grains with distinct orientations, with the coefficients $\underline{W}_\sf(T)$ and $\underline{W}_\gb(T)$ related to the barrier heights among minima, varying with temperature \cite{yang2019npj, yang2023npj}. The heat contribution $f_\text{ht}$ thereby manifests the stability of states by shifting the minima according to the local temperature field. 
$c_\mathrm{r}$ is a relative specific heat landscape formulated by the volumetric specific heats in the substance 
$c_\subst$ and pore $c_\at$, i.e., $c_\mathrm{r}=h_{\subst} c_\subst + h_\at c_\at$ with corresponding interpolation functions 
$h_\subst$, $ h_\at$ indicating the substance (including solid and liquid) and the atmosphere/pore, respectively. 
When the material reached melting point $T_\mathrm{M}$, the extra contribution due to the latent heat $\mathcal{L}_\mathrm{M}$ is mapped by the interpolation function $h_\text{M}$, which approaches unity when $T \rightarrow T_\text{M}$ \cite{yang2019npj, yang2023npj}. It is worth noting that $\underline{W}_\sf(T)$, $\underline{\kappa}_\sf(T)$, $\underline{W}_\gb(T)$, and $\underline{\kappa}_\gb(T)$ inherit their temperature dependencies from the surface energy $\gamma_\sf(T)$ and grain boundary energy $\gamma_\gb(T)$, respectively (see Sec. \ref{sec:para}). At the equilibrium, they can be related to $\gamma_\sf(T)$, $\gamma_\gb(T)$, and a diffuse-interface width of the grain boundary $\lgb$, which have been explained in our former work \cite{yang2019npj, yang2023npj}.
% \begin{equation}
% \begin{split}
%  &\Gamma_{\sf}(T)=\frac{\sqrt{2}}{6}\tau_{\sf}(T)\sqrt{({W}_\sf+7W_\gb)(\kappa_\sf+\kappa_\gb)}, 
%  \\
%  &\Gamma_{\gb}(T)=\frac{2\sqrt{3}}{3}\tau_{\gb}(T)\sqrt{{W}_\gb\kappa_\gb}, \\
%  &\ell_{\gb}\approx\frac{2\sqrt{3}}{3}\sqrt{\frac{\kappa_\gb}{{W}_\gb}},
% \end{split}
% \end{equation}

The Cahn-Hilliard mobility employed in \eqref{eqk1} equation adopts the anisotropic form. Contributions considered in this work contain not only the mass transfer through substance, atmosphere, surface and grain boundary, but also the diffusion enhancement due to possible partial melting \cite{yang2020investigation, yang2023npj, oyedeji2023variational}, i.e.
\begin{equation}
\begin{split}
\mathbf{M}=\frac{1}{2(\underline{W}_\sf+\underline{W}_\gb)}\left[h_\subst D_\subst\mathbf{I} + h_\at D_\at\mathbf{I} + h_\sf D_\sf\mathbf{T}_\sf+h_\gb D_\gb \mathbf{T}_\gb\right]
+h_\mathrm{M}M_\mathrm{M}\mathbf{T}_\sf,
\label{eqk2}
\end{split}
\end{equation}
where $D_\mathrm{path}$ is the effective diffusivity of the mass transport via $\mathrm{path}=\mathrm{ss},~\mathrm{at},~\mathrm{sf},~\mathrm{gb}$, and $ h_\text{ss}$, $ h_\text{at}$, $ h_\text{sf}$ and $ h_\text{gb}$ are again interpolation functions to indicate substance, atmosphere/pore, surface and grain boundary, respectively, which obtain unity only in the corresponding region \cite{yang2020investigation, yang2023npj, oyedeji2023variational}. Mass transport along surface and grain boundary is also regulated by corresponding projection tensor $\mathbf{T}_\gb$ and $\mathbf{T}_\sf$. 
It is worth noting that the partial melting contribution $M_\text{M}$ is treated as an enhanced surface diffusion when $T \rightarrow T_\text{M}$ due to the assumption of a limited melting phenomenon around the surface of particles. In this sense, formulations in Eqs. (\ref{eqk1}) and (\ref{eqk2}) disregard the contributions from melt flow dynamics as well as from inter-coupling effects between mass and heat transfer, i.e., Soret and Dufour effects \cite{yang2020investigation}. 
The isotropic Allen-Cahn mobility is explicitly formulated by the the temperature-dependent GB mobility $G_\text{gb}$, GB energy $\gamma_\text{gb}$ and gradient coefficient $\underline{\kappa}_\eta$ as \cite{moelans2008quantitative, turnbull1951theory}
\begin{equation}
L=\frac{G_\text{gb}\gamma_\text{gb}}{\underline{\kappa}_\eta}.
\label{eqk4}
\end{equation} 
The phase-dependent thermal conductivity tensor takes the continuity of the thermal flux along both the normal and tangential directions of the surface into account, and is formulated as \cite{yang2022diffuse, nicoli2011tensorial} 
\begin{equation}
 \mathbf{K}=\left[h_\subst K_\subst + h_\at K_\at\right]\mathbf{T}_\sf+\left[\frac{K_\subst K_\at}{h_\subst K_\at + h_\at K_\subst}\right]\mathbf{N}_\sf,
\label{eq24}
\end{equation}
where $K_\subst$ and $K_\at$ are the thermal conductivity of the substance and pore/atmosphere. $\mathbf{N}_\sf$ is the normal tensor of the surface \cite{yang2022diffuse, yang2023npj, oyedeji2023variational}. It is worth noting that radiation contribution via pore/atmosphere is considered in $K_\at$ as $ K_\at = K_0 + 4T^3\sigma_\mathrm{B}\ell_\mathrm{rad}/3$ with $K_0$ the thermal conductivity of the Argon gas and $\sigma_\mathrm{B}$ the Stefan-Boltzmann constant. $\ell_\mathrm{rad}$ is the effective radiation path between particles, which usually takes the average diameter of the powders \cite{denlinger2016thermal}.
% {The above interpolation, however, neglects the heat resistance due to ... on the surface, which should be modeled using the anisotropic thermal conduction with the direction-specified interpolation. This will be separately discussed in our future works.}

The beam-incited thermal effect is equivalently treated as a volumetric heat source with its distribution along the depth direction formulated in a radiation penetration fashion, as the powder bed is regarded as an effective homogenized optical medium, i.e.
\begin{equation*}
q(\vec{r},t)=Pp_{xy}[\vec{r}_O(\vvec,t)]\frac{\text{d}a}{\text{d}z}
 \label{qexpr},
\end{equation*}
in which the in-plane Gaussian distribution $p_{xy}$ with a moving center $\vec{r}_O(\vvec,t)$. $P$ is the beam power and $\vvec$ is the scan velocity with its magnitude $v=|\vvec|$ as the scan speed, which are two major processing parameters in this work. The absorptivity profile function along depth ${\text{d}a}/{\text{d}z}$ which is calculated based on Refs. \cite{yang2019npj,Gusarov2009}. 
% \begin{equation}
%  \begin{split}
%  \mathcal{L}_\rho &= \frac{\partial e_\mathrm{pt}}{\partial \rho} = 2\underline{C}_\mathrm{pt}\rho(1-3\rho+2\rho^2) + 2\underline{D}_\mathrm{pt}\left[ \rho - 3\sum_i\eta_i^2 + 2\sum_i\eta_i^3\right],\\
%  \mathcal{L}_{\eta_i} &= \frac{\partial e_\mathrm{pt}}{\partial \eta_i} =12\underline{D}_\mathrm{pt}\eta_i\left[1-\rho+\eta_i\left(\rho-2\right)+\sum_i\eta_i^2
%  \right]
% \label{eq:lh}
%  \end{split}
% \end{equation}
It is obvious that when one ignores the effects of those latent heat induced by microstructure evolution (i.e. the evolution of the pore/substance as well as unique grains), Eq. (\ref{eqk5}) can be degenerated to the conventional Fourier’s equation for heat conduction with an internal heat source.

\subsection{Elasto-plastic model for thermo-mechanical analysis}\label{sec:tep}

The transient temperature field and substance field $\rho$, resulting from the non-isothermal phase-field simulations, are then imported into the quasi-static elasto-plastic model as the thermal load as well as the phase indicator for the interpolation of mechanical properties. The linear momentum equation reads
\begin{equation}
 \nabla\cdot\boldsymbol{\sigma}=\vec{0},
 \label{eq:gov_mech}
\end{equation}
where $\boldsymbol{\sigma}$ is the $2^\mathrm{nd}$-order stress tensor. Taking the Voigt-Taylor interpolation scheme (VTS), where the total strain is assumed to be identical among pores/atmosphere and substance, i.e. $\boldsymbol{\varepsilon}= \boldsymbol{\varepsilon}_\mathrm{ss}=\boldsymbol{\varepsilon}_\mathrm{at}$.~\cite{voigt1889, Schneider2015, durga2015}. In this regard, the stress can be eventually formulated by the linear constitutive equation
\begin{equation}
 \boldsymbol{\sigma} = \Cten(\rho, T):\left(\boldsymbol{\varepsilon}-
 \boldsymbol{\varepsilon}^\mathrm{th} - \boldsymbol{\varepsilon}^\mathrm{pl}\right)
 \label{eq:cons_eq}
\end{equation}
where the $4^\mathrm{th}$-order elastic tensor is interpolated from the substance one $\Cten_\mathrm{ss}$ and the pores/atmosphere one $\Cten_\mathrm{at}$, i.e.,
\begin{equation}
 \Cten(\rho, T) =h_\mathrm{ss}(\rho)\Cten_\mathrm{ss}(T) +h_\mathrm{at}(\rho)\Cten_\mathrm{at}.
 \label{eq:C}
\end{equation}
In this work, $\Cten_\mathrm{ss}$ is calculated from the temperature-dependent Youngs' modulus $E(T)$ and poison ratio $\nu(T)$, while $\Cten_\mathrm{at}$ is assigned with a sufficiently small value. 

The thermal eigenstrain $\boldsymbol{\varepsilon}^\mathrm{th}$ is calculated using interpolated coefficient of thermal expansion, i.e.,
\begin{equation}
\begin{split}
 \boldsymbol{\varepsilon}^\mathrm{th}&=\alpha(\rho, T)[T-T_0]\mathbf{I},\\
 \alpha(\rho, T) &=h_\mathrm{ss}(\rho)\alpha_\mathrm{ss}(T)+h_\mathrm{at}(\rho)\alpha_\mathrm{at}.
\end{split}
\end{equation}
Here $\mathbf{I}$ is the $2^\mathrm{nd}$-order identity tensor. Meanwhile, for plastic strain $\boldsymbol{\varepsilon}_\mathrm{pl}$ and isotropic hardening model with the von Mises yield criterion is employed. 
The yield condition is determined as
\begin{equation}
 f(\boldsymbol{\sigma}, p_\mathrm{e} ) = \sigma_\mathrm{e} - [{\sigma}_\mathrm{y}(T) + \Hm(T) p_\mathrm{e} ]\leq 0 
 \label{eq:yieldcon}
\end{equation}
with
\begin{equation*}
\sigma_\mathrm{e}=\sqrt{\frac{3}{2}\mathbf{s}:\mathbf{s}},\quad p_\mathrm{e}=\int
\sqrt{\frac{2}{3} \mathrm{~d} \boldsymbol{\varepsilon}^{\mathrm{pl}}:\mathrm{d} \boldsymbol{\varepsilon}^{\mathrm{pl}}},
\end{equation*}
where $\sigma_\mathrm{e}$ is the von Mises stress. $\mathbf{s}$ is the deviatoric stress, and ${\sigma}_\mathrm{{y}}$ is the isotropic yield stress when no plastic strain is present. The isotropic plastic modulus $\Hm$ can be calculated from the isotropic tangent modulus $E_\mathrm{t}$ and the Young's modulus $E$ as $\Hm = EE_\mathrm{t}/(E-E_\mathrm{t})$. 
$p_\mathrm{e} $ is the effective (accumulated) plastic strain, which is integrated implicitly by the plastic strain increment $\boldsymbol{\varepsilon}^{\mathrm{pl}}$ employing radial return method and will be elaborated in \secref{sec:ni}.

% \section{Simulation setup}

\subsection{Boundary conditions and model parameters} \label{sec:para}

Together with the governing equations shown in \eqsref{eqk1}{eqk5}, the following BCs are also employed for the process simulations
\begin{empheq}[left=\empheqlbrace]{align}
&\left.\nabla \rho \right|_\Gamma \cdot \nvec = 0
,\label{eq:bc1}\\
&\left.\mathbf{K}\cdot\nabla T \right|_{\Gamma_\mathrm{S}\cup\Gamma_\mathrm{T}} \cdot \nvec =h_\mathrm{at} \left[ \underline{h}\left(\left.T\right|_{\Gamma_\mathrm{S}\cup\Gamma_\mathrm{T}}-T_{0}\right)+\varepsilon \sigma_{\mathrm{B}}\left(\left.T\right|_{\Gamma_\mathrm{S}\cup\Gamma_\mathrm{T}} ^{4}-T_{0}^{4}\right)\right]
,\label{eq:bc2}\\
&\left.T\right|_{\Gamma_{\mathrm{B}}}=T_{0}\label{eq:bc3}
\end{empheq}
with the convectivity $\underline{h}$, Stefan-Boltzmann constant $\sigma_{\mathrm{B}}$, the hemispherical emissivity $\varepsilon$, and the pre-heating (environmental) temperature $T_{0}$. $\Gamma_\mathrm{T}$ and $\Gamma_\mathrm{B}$ are correspondingly the top and bottom boundaries of the simulation domain, and $\Gamma_\mathrm{S}$ is the set of all surrounded boundaries. $\Gamma_\mathrm{B}\cup\Gamma_\mathrm{S}\cup\Gamma_\mathrm{T}=\Gamma$, as the schematic shown in the inset of \figref{fig:wf}b. $\nvec$ is the normal vector of the boundary. \eqref{eq:bc1} physically shows the close condition for the mass transfer, restricting no net mass exchange with the environment. \eqref{eq:bc2} shows the heat convection and radiation, which are only allowed via the pore/atmosphere at the boundary (masked by $ h_\mathrm{at}$). \eqref{eq:bc3} emulates a semi-infinite heat reservoir under the bottom of the substrate with constant temperature $T_0$, consistently draining heat from the simulation domain and reducing its temperature back to $T_0$. 

As summarized in Sec. \ref{sec:npt}, this simulation requests following parameters/properties: the thermodynamic parameters $\uW{sf}$, $\uW{gb}$, $\ukappa{\rho}$, and $\ukappa{\eta}$; the kinetic properties (diffusivities and GB mobility) $D_\mathrm{path}$ (path = ss, at, sf, gb) and $G_\gb$; and the thermal properties $K_\mathrm{ss}$ and $K_\mathrm{at}$. Among them, $\uW{sf}$, $\uW{gb}$, $\ukappa{\rho}$, and $\ukappa{\eta}$ are parameterized by the temperature-dependent surface as well as GB energies, and a nominal diffuse interface width $\lgb$ as
\begin{equation}
\begin{split}
 &\gamma_{\sf}(T)=\frac{\sqrt{2}}{6}\tau_{\sf}(T)\sqrt{({W}_\sf+7W_\gb)(\kappa_\rho+\kappa_\eta)}, 
 \\
 &\gamma_{\gb}(T)=\frac{2\sqrt{3}}{3}\tau_{\gb}(T)\sqrt{{W}_\gb\kappa_\eta}, \\
 &\lgb\approx\frac{2\sqrt{3}}{3}\sqrt{\frac{\kappa_\eta}{{W}_\gb}}
\end{split}
\end{equation}
with normalized tendencies $\tau_{\sf}(T)$ and $\tau_{\gb}(T)$ that reach unity at $\Tm$.  $\uW{sf} \approx {W}_\sf\tau_{\sf}(T)$, $\uW{gb} = {W}_\gb\tau_{\gb}(T)$, $\ukappa{\rho} \approx \kappa_\rho\tau_{\sf}(T)$, and $\ukappa{\eta} = \kappa_\eta\tau_{\gb}(T)$. Constants ${W}_\sf$, $W_\gb$, $\kappa_\rho$, $\kappa_\eta$ also satisfy a constraint $({W}_\sf+{W}_\gb)/\kappa_\rho = 6W_\gb/\kappa_\eta$ derived from the coherent diffuse-interface profile at equilibrium (Supplementary Note 1 of Ref. \cite{yang2019npj}). These parameter/properties are collectively shown in \tabref{tab1}. 

For the thermo-elasto-plastic calculations, a $250\times250\times250$ \si{\micro m^3} subdomain was selected from the center of the prime domain to eliminate the boundary effects (\figref{fig:wf}a). The above momentum balance is subjected to the following rigid support BC as
\begin{equation}
 \left.\uvec\right|_{\Gamma_\mathrm{S}\cup\Gamma_\mathrm{B}}\cdot\nvec=0,
\end{equation}
restricting the displacement $\uvec$ along the normal direction of all boundaries except the top ($\Gamma_\mathrm{T}$), which is traction free. As summarized in Sec. \ref{sec:tep}, temperature-dependent $E$, $\nu$, $\alpha$, $E_\mathrm{t}$, and $\ystress$ are as listed in \tabref{tab:mech}, where piecewise linear interpolation was employed to implement their temperature dependence.

\subsection{Numerical implementation and parallel computing}\label{sec:ni}
The theoretical models are numerically implemented via the finite element method within the program NIsoS, developed by authors based on MOOSE framework (Idaho National Laboratory, ID, USA) \cite{permann2020moose}. 8-node hexahedron Lagrangian elements are chosen to mesh the geometry. A transient solver with preconditioned Jacobian-Free Newton-Krylov method (PJFNK) and backward Euler algorithm was employed for both problems.

For non-isothermal phase-field simulations, the Cahn–Hilliard equation in \eqref{eqk1} was solved in a split way, The constraint of the order parameters was fulfilled using the penalty method. To reduce computation costs, h-adaptive meshing and time-stepping schemes are used. The additive Schwarz method (ASM) preconditioner with incomplete LU-decomposition sub-preconditioner was also employed for parallel computation of the vast linear system, seeking the balance between memory consumption per core and computation speed \cite{balay2019petsc}. Due to the usage of adaptive meshes, the computational costs vary from case to case. The peak DOF number is on the order of 10,000,000 for both the nonlinear system and the auxiliary system. The peak computational consumption is on the order of 10,000 core-hour. More details about the FEM implementation are shown in the Supplementary Note 4 of Ref. \cite{yang2019npj}.

For thermo-elasto-plastic simulations, a static mesh was utilized to avoid the hanging nodes generated from h-adaptive meshing scheme. In that sense, the transient fields $T$ and $\rho$ of each calculation step were uni-directionally mapped from the non-isothermal phase-field results (with h-adaptive meshes) into the static meshes, assuming a weak coupling between thermo-structural and mechanical problems in this work. This is achieved by the MOOSE-embedded \texttt{SolutionUserObject} class and associated functions. The parallel algebraic multigrid preconditioner BoomerAMG was utilized, where the Eisenstat-Walker (EW) method was employed for determining linear system convergence. It is worth noting that a vibrating residual of non-linear iterations would show without the employment of EW method for this work. The DOF number of each simulation is on the order of 1,000,000 for the nonlinear system and 10,000,000 for the auxiliary system. The computational consumption is approximately 1,000 CPU core-hour. 

A modified radial return method was employed to calculate the integral of the plasticity as well as to determine the yield condition during the process with the temperature-dependent elasto-plastic properties at any time $t$ with the time increment $\Delta t$. This method employs the trial stress $\stress^\star$ calculated assuming an elastic new strain increment $\Delta \strain^\star$
\begin{equation}
 \stress^\star = \Cten(T_{t}):\left[\strain^\mathrm{el}_{t}+\Delta\strain^\star\right],
 \label{eq:trialstress}
\end{equation}
where the elasticity tensor is obtained under the current temperature field $T_{t}$. 
Once the trial stress state is outside the yield condition (\eqref{eq:yieldcon}), i.e., the plastic flow exists, the stress is then projected onto the closet point of the expanded yield surface with the normal direction determined as $\mathbf{n}_\mathrm{y}={3\stressdev^\star}/{2\vm^\star}$ with the von Mises $\vm^\star$ and deviatoric trial stress $\stressdev^\star$ \cite{dunne2005introduction, simo2006computational} . Meanwhile, assuming isotropic linear hardening under $T_{t}$ of every timestep, the amount of the effective plastic strain increment $\Delta p$ for returning the stress state back to the yield surface is calculated in an iterative fashion adopting Newton's method 
\begin{gather}
 \diff \Delta p = \frac{\vm^\star -3 G(T_{t}) \Delta p_{t}- \left[\Hm(T_{t}) p_{t} + \ystress(T_{t}) \right]}{3G(T_{t}) + \Hm(T_{t})},\label{eq:rrm1}\\
 \Delta p_{(t+\Delta t)} =\Delta p_{t} + \diff \Delta p ,\label{eq:rrm2}\\
 p_{(t+\Delta t)} = p_{t} + \Delta p_{(t+\Delta t)}
\end{gather}
where $p_{t}$ and $\Delta p_{t}$ are updated as $p_{(t+\Delta t)}$ and $\Delta p_{(t+\Delta t)}$ at the end of the timestep ($t+\Delta t$). $G(T_{t})$ and $\Hm(T_{t})$ are shear and isotropic plastic modulus calculated under the local temperature at the current timestep ($T_{t}$). Here $G(T_{t})=E(T_{t})/[2 + 2\nu(T_{t})]$ with $E(T_{t})$ and $\nu(T_{t})$ correspondingly the temperature-dependent Young's modulus and Poisson ratio.
Knowing that the plastic strain increment $\Delta\strain^\mathrm{pl}=\Delta p\mathbf{n}_\mathrm{y}$ following the normality hypothesis of plasticity \cite{dunne2005introduction}, the updated stress and plastic strain at the end of timestep are thereby obtained by
\begin{equation}
\begin{split}
\stress_{(t+\Delta t)} &=\left\{ 
\begin{aligned}
 &\stress^\star - 2G(T_{t})\Delta p_{(t+\Delta t)} \mathbf{n}_\mathrm{y}, \quad& T<T_\mathrm{A},\\
 &\mathbf{0},\quad& T\geq T_\mathrm{A},
\end{aligned}
\right.\\
\strain^\mathrm{pl}_{(t+\Delta t)} &=
\left\{ 
\begin{aligned}
 &\strain^\mathrm{pl}_{t} + \Delta p_{(t+\Delta t)}\mathbf{n}_\mathrm{y}, \quad& T<T_\mathrm{A},\\
 &\mathbf{0},\quad& T\geq T_\mathrm{A},
\end{aligned}
\right.
\label{eq:stressstrainupdate}
\end{split}
\end{equation}
in which the vanishing of the stress as well as the plastic strain beyond a stress-free temperature $T_\mathrm{A}$ (in this work $T_\mathrm{A} = \Tm$) is also considered. 

\eqsref{eq:trialstress}{eq:stressstrainupdate} are sequentially executed on every 
timestep and update the quantities under $T_{t}$. 
It is worth noting that the linear-interpolated $\Hm$ is implemented for both $p$- and $T$-dependence
\begin{equation}
 \Hm(p_t, T_t)=(1-\tau_t)f(\hat{T}_i, p_t)+\tau_t f(\hat{T}_{i+1}, p_t)
\end{equation}
with
\begin{align*}
 \tau_t=\frac{T_t - \hat{T}_i}{\hat{T}_{i+1}-\hat{T}_i},
\end{align*}
where $f(\hat{T}_i, p_t)$ is a piecewise function with the grids $\hat{T}_i$ ($i=1,2,...$) and $T_i$ is inside the section bound by $\hat{T}_i$ and $\hat{T}_{i+1}$, i.e., $T_t\in (\hat{T}_i,\hat{T}_{i+1}]$. Considering reduction of the non-linearity, the $p$-independent $\Hm(T)$ was practically employed in the simulations, as formulated in \eqref{eq:rrm1}.

% Meanwhile, to calculate the amount of the effective plastic strain increment $\Delta p$ for returning the stress state back to yield surface, the following relation (with detailed derivation shown in Supplementary Note 1) is considered
% \begin{subequations}
% \begin{equation}
%  \vm_{(t+\Delta t)} &= \sigma^\star - 3G(T_{t})\Delta p\label{eq:rrm1}
% \end{equation}
% with the von Mises updated stress $\vm_{(t+\Delta t)}$ determined by yield condition under $T_{t}$, i.e.,
% \begin{equation}
%  \vm_{(t+\Delta t)} &- \left[H(T_{t}) p_{t} + \ystress(T_{t}) \right] = 0,\label{eq:rrm2}
% \end{equation}
% \end{subequations}
% where $G(T_{t})$ and $H(T_{t})$ are shear and isotropic plastic modulus under $T_{t}$

\section*{Data Availability}
The authors declare that the data supporting the findings of this study are available within the paper. The temperature-dependent parameters, thermodynamic database, simulation results, supplementary data and utilities are cured in the online dataset (DOI: \url{10.5281/zenodo.xxxxxxx}).

\section*{Code Availability}
Source codes of MOOSE-based application NIsoS and related utilities are available and can be accessed via the online repository \url{bitbucket.org/mfm_tuda/nisos.git}.

\section*{Acknowledgements}
Authors acknowledge the financial support of German Research Foundation (DFG) in the framework of the Collaborative Research Centre Transregio 270 (CRC-TRR 270, project number 405553726, sub-projects A06 and A07), the Research Training Groups 2561 (GRK 2561, project number 413956820, sub-project A4), the Priority Program 2122 (SPP 2122, project number 493889809). X. Zhou acknowledges the support from the National Natural Science Foundation of China (project number 12302231), Sichuan Science and Technology Program (project number 2023NSFSC0910), and China Postdoctoral Science Foundation (project number 2023M732433). The authors also greatly appreciate the access to the Lichtenberg II High-Performance Computer (HPC) and the technique supports from the HHLR, Technische Universit\"at Darmstadt. The computating time on the HPC is granted by the NHR4CES Resource Allocation Board under the project ``special00007''. Y. Yang and S. Bharech also greatly thanks Dr. B. Lin for helping with the setup of layerwise powder deposition, and P. Kühn and J. Ma for their extensive help with language and grammar checks.

\section*{Competing Interests}
The authors declare no competing financial or non-financial interests.

\section*{Author Contributions}
Conceptualization: B.-X.X., X.Z. and Y.Y.; methodology: X.Z., Y.Y., J.S. and B.-X.X.; software: Y.Y. and X.Z.; investigation: Y.Y., S.B., and N.F.; formal analysis: Y.Y., S.B., and N.F.; resources, X.Z., J.S. and B.-X.X.; data curation: Y.Y.; writing—original draft preparation: Y.Y., S.B., N.F., X.Z. and B.-X.X.; writing—review and editing: Y.Y., S.B., N.F., X.Z., J.S. and B.-X.X.; visualization: Y.Y. and S.B.; supervision: B.-X.X.; consultation and discussion: J.S.; funding acquisition: B.-X.X. All authors have read and agreed to the published version of the manuscript.

%\nocite{*}% Show all bib entries - both cited and uncited; comment this line to view only cited bib entries;
%\clearpage

%%%%%%%%%%%%%%%%%%%%%%%%%%%%%%%%%%%%%%
%% Tables
\clearpage

\begin{table}[!t]
	\centering 
	\begin{threeparttable}
		\caption{Material properties of the bulk SS316L used in the non-isothermal phase-field simulations. Here $\mathfrak{R}$ represents the ideal gas constant.}
		\begin{tabular}{ccccccc}
			\hline
			Properties & Expressions ($T$ in K) & Units &References \\ \hline
			$T_\text{M}$ & $\sim1700$ & K & & \\
 $\gamma_{\mathrm{sf}}$ & $10.315 -5.00 \E{-3}T$& J/m$^2$& \cite{price1964surface} \\
 $\gamma_{\mathrm{gb}}$ & $13.018 -7.50 \times \E{-3}T$ & J/m$^2$& \cite{price1964surface} \\
 $\lgb$ & $2\E{-6}$ & m & &\\
			$D_\text{sf}$ & $0.40\text{exp}\left(-2.200\times 10^5/\mathfrak{R}T \right)$ & m$^2$/s&\cite{blakely1963studies} \\
			$D_\text{gb}$ & $2.40\times 10^{-3}\text{exp}\left(-1.770\times 10^5/\mathfrak{R}T \right)$&m$^2$/s & \cite{blakely1963studies} \\
			$D_\text{ss}$ & $2.17\times 10^{-5}\text{exp}\left(-2.717\times 10^5/\mathfrak{R}T \right)$&m$^2$/s & \cite{mead1956self} \\
			$G_\text{gb}$ & $3.26\times 10^{-3}\text{exp}\left(-1.690\times 10^5/\mathfrak{R}T \right)$ \tnote{*} & m$^4$/(J s) &\\
			$M_\text{M}$ & $\sim3.45\times 10^{-13}$ \tnote{**} & m$^5$/(J s) &\\
			$K_\text{ss}$ & $10.292+0.014T$ & J/(s m K) & \cite{liu2012micro} \\
			$K_\text{at}$ & $\sim0.06$ & J/(s m K) & \cite{hoshino1986determination} \\
			$c_\text{ss}$ & $3.61\times 10^6+1272T$ & J/(m$^3$ \text{K}) & \cite{liu2012micro} \\
			$c_\text{at}$ & 717.6 & J/(m$^3$ \text{K}) & \cite{chase1998nist}\\
			$\mathcal{L}_\text{M}$ & $2.4\times 10^9$ & J/m$^3$ & \cite{liu2012micro} \\
			\hline
		\end{tabular}
		\label{tab1}
		\begin{tablenotes}
			\footnotesize
			\item[*] Activation energy is obtained from \cite{di2002analysis} while the prefix factor is estimated as unity at $T_\text{M}$ after normalization. 
			\item[**] Estimated as $100D_\text{sf}/2(\uW{sf}+\uW{gb})$.
		\end{tablenotes}
	\end{threeparttable}
\end{table}

%%% tab:mech
\begin{table}[!h]
	\centering 
 	\begin{threeparttable}
 \caption{Temperature-dependent mechanical properties of the bulk SS316L used in the thermo-elasto-plastic simulations \cite{mills2002recommended, li2018modeling}}
\begin{tabular}{cccccccc}
\hline
$T$ (K) & 298 & 873 & 1073 & 1473 & 1623 & $\geq 1773$ \\ \hline
$\nu$ & 0.33 & 0.35 & 0.36 & 0.38 & 0.39 & 0.40\\
$E$ (MPa) & $2.00\E5$ & $1.35\E5$ & $7.75\E4$ & $1.21\E4$ & $6.14\E3$ & $200$ \\
$\sigma_\mathrm{y}$ (MPa) & $3.45\E2$ & $2.12\E2$ & $1.99\E2$ & $100$ & $50$ & 5 \\
$E_\mathrm{t}$ (MPa) & $5.89\E3$ & $1.70\E3$ & $1.40\E3$ & $100$ & $10$ & $1$ 
\\
$\Hm$ (MPa) & $6.07\E3$ & $1.72\E3$ & $1.43\E3$ & $101$ & 10 & 1 \\ 
$\alpha$ (1/K) & $1.20\E{-5}$ & $1.30\E{-5}$ & $1.32\E{-5}$ & $1.36\E{-5}$ & $1.38\E{-5}$ & $1.40\E{-5}$ \\
\hline
\end{tabular}
\label{tab:mech}
\end{threeparttable}
\end{table}

% %%% tab:reg
% \begin{table}[!h]
% 	\centering 
%	\begin{threeparttable}
%\caption{Regression analyses on scaling relation of $\bar{p}_\mathrm{e}^\mathrm{pb}(\varrho)$ and $\bar{\sigma}_\mathrm{e}^\mathrm{pb}(U_\mathrm{e})$.}
% 		\begin{tabular}{cccccc}
% \hline
% Relation & $C$ & $\varsigma_\mathrm{C}$ & $I$ &$\varsigma_\mathrm{I}$ & $R^2$ (\%) \\ \hline
% $\bar{p}_\mathrm{e}^\mathrm{pb}(\varrho)$ & $5.20\E{-2}$ & $4.62\E{-3}$ & 1.73 & 0.12 & 98.05 \\
% $\bar{\sigma}_\mathrm{e}^\mathrm{pb}(U_\mathrm{L})$ \tnote{*}& $6.52\E2$& $8.53\E2$ & 0.51& $5.41\E{-2}$ & 91.38 \\
% 			\hline
% 		\end{tabular}
% 		\label{tab:reg}
%  \begin{tablenotes}
% \footnotesize
% \item[*] The unit of $\bar{\sigma}^\mathrm{pb}_\mathrm{e}$ and $U_\mathrm{e}$ are Pa or \si{J~m^{-3}}.
% 		\end{tablenotes}
% 	\end{threeparttable}
% \end{table}

%%%%%%%%%%%%%%%%%%%%%%%%%%%%%%%%%%%%%%%%
%% Figures
\clearpage

% \begin{figure}[!h]
% 	\centering
%	\includegraphics[width=18cm]{fig_poro_vis.pdf}
% 	\caption{\textcolor{red}{(a$_{1-9}$)(c) Comparison of the proposed relation between densification of the microstructure (represented by a so-called densification factor, $\varrho$) and specific energy input of the beam ($\uv$) with the simulation results and existing research works.}}
% 	\label{fig:pmap}
% \end{figure}

% \begin{figure}[!h]
% 	\centering
%	\includegraphics[width=18cm]{fig_pe-vM_dist.pdf}
% 	\caption{\textcolor{red}{Local distribution of plastic strain and von Mises stress in the RVE after the multilayer SLS process simulation with (a$_{1-5}$) varying beam power and scan speed.}}
% 	\label{fig:vm}
% \end{figure}

% \begin{figure}[!h]
% 	\centering
%	\includegraphics[width=18cm]{fig_clip_avg_sig.pdf}
% 	\caption{xxx.}
% 	\label{fig:vm}
% \end{figure}

% \begin{figure}[!t]  
% 	\centering
%	\includegraphics[width=9cm]{fig_poro_vm_pe.pdf}
% 	\caption{Power law regressions of (a) residual von Mises stress $\barvm^\mathrm{S}$ and (b) effective plastic strain $\barpe^\mathrm{S}$ in strut average on porosity.} 
% 	\label{fig:pvp}
% \end{figure}

\clearpage
% \bibliographystyle{naturemag}
% \bibliography{reference, temp_ref}%

\end{document}